\documentclass{llncs}
\usepackage{epsfig}
\usepackage{multicol} 
\usepackage{makeidx}  

\begin{document}
\frontmatter          
\pagestyle{headings}  
\addtocmark{Pedestrian, Crowd, and Evacuation Dynamics} 
\mainmatter              
\titlerunning{Pedestrian, Crowd, and Evacuation Dynamics}
\title{Pedestrian, Crowd, and Evacuation Dynamics}
\author{Dirk Helbing\inst{1,2} and Anders Johansson\inst{1}}
\authorrunning{D. Helbing and A. Johansson}   
%
\tocauthor{Dirk Helbing (TU Dresden)}
\tocauthor{Anders Johansson (TU Dresden)}
\institute{Dresden University of Technology, Andreas-Schubert-Str. 23, 01062 Dresden, Germany
\and
Collegium Budapest~-- Institute for Advanced Study, Szenth\'{a}roms\'{a}g utca 2, 1014 Budapest, Hungary}

\maketitle              

\section{Article Outline}

This contribution describes efforts to model the behavior of individual pedestrians and their 
interactions in crowds, which generate certain kinds of self-organized patterns of motion. Moreover, this article
focusses on the dynamics of crowds in panic or evacuation situations, methods to optimize building
designs for egress, and factors potentially causing the breakdown of orderly motion.

\section{Glossary}

{\noindent\bf Collective Intelligence:} Emergent functional behavior of a large number of people that results from 
{\em interactions} of individuals rather than from individual reasoning or global optimization.

{\noindent\bf Crowd:} Agglomeration of many people in the same area at the same time. The density
of the crowd is assumed to be high enough to cause continuous interactions with or reactions to other individuals.

{\noindent\bf Crowd Turbulence:} Unanticipated and unintended irregular motion of individuals into different directions due to
strong and rapidly changing forces in crowds of extreme density.

{\noindent\bf Emergence:} Spontaneous establishment of a qualitatively new behavior through non-linear interactions
of many objects or subjects.

{\noindent\bf Evolutionary Optimization:} Gradual optimization based on the effect of frequently repeated
random mutations and selection processes based on some success function (``fitness''). 

{\noindent\bf Faster-is-Slower Effect:} This term reflects the observation that certain processes (in evacuation situations,
production, traffic dynamics, or logistics) take more time if performed at high speed. In other words, waiting can often 
help to coordinate the activities of several competing units and to speed up the average progress.

{\noindent\bf Freezing-by-Heating Effect:} Noise-induced blockage effect caused by the 
breakdown of direction-segregated walking patterns (typically two or more ``lanes'' characterized 
by a uniform direction of motion). ``Noise'' means frequent variations of the walking direction due to
nervousness or impatience in the crowd, e.g. also frequent overtaking maneuvers in dense, slowly moving
crowds.

{\noindent\bf Panic:} Breakdown of ordered, cooperative behavior of individuals due to anxious reactions to a certain event. Often, panic is characterized by attempted escape of many individuals 
from a real or perceived threat in situations of imminent danger,
which may end up in trampling or crushing of people in a crowd. 
Definitions of the term ``panic'' are controversial and depend on the discipline or community of people using it.

{\noindent\bf Self-Organization:} Spontaneous organization (i.e. formation of ordered patterns) not induced by
initial or boundary conditions, by regulations or constraints. Self-organization is a result of non-linear interactions
between many objects or subjects, and it often causes different kinds of spatio-temporal patterns of motion.

{\noindent\bf Social Force:} Vector describing acceleration or deceleration effects 
that are caused by social interactions rather than by physical interactions or 
fields.

\section{Definition}
The modeling of pedestrian motion is of great theoretical and practical interest. Recent experimental efforts
have revealed quantitative details of pedestrian interactions, which have been successfully
cast into mathematical equations. Furthermore, corresponding computer simulations of large numbers of 
pedestrians have been compared with the empirically observed dynamics of crowds. Such 
studies have led to a deeper understanding of how collective behavior on a macroscopic 
scale emerges from individual human interactions. Interestingly enough, the non-linear interactions of pedestrians
lead to various complex, spatio-temporal pattern-formation phenomena. This includes the emergence 
of lanes of uniform walking direction, oscillations of the pedestrian flow at bottlenecks, and the formation
of stripes in two intersecting flows. Such self-organized patterns of motion demonstrate that an efficient, 
``intelligent'' collective dynamics can be based on simple, local interactions. 
Under extreme conditions, however,
coordination may break down, giving rise to critical crowd conditions. Examples are 
``freezing-by-heating'' and ``faster-is-slower'' effects, but also the transition to ``turbulent'' crowd dynamics. 
These observations have important implications for the optimization of pedestrian facilities, in particular for
evacuation situations.

\section{Introduction}

The emergence of new, functional or complex collective behaviors in social systems has fascinated
many scientists. One of the primary questions in this field is how
cooperation or coordination patterns originate based on elementary individual interactions. 
While one could think that these are a result of intelligent human actions, it turns out that
much simpler models assuming automatic responses can reproduce the observations very well.
This suggests that humans are using their intelligence primarily for more complicated
tasks, but also that simple interactions can lead to intelligent patterns of motion. Of course, it is reasonable
to assume that these interactions are the result of a previous learning process that has optimized the
automatic response in terms of minimizing collisions and delays. This, however, seems to be sufficient
to explain most observations.
\par
Note, however, that
research into pedestrian and crowd behavior is highly multi-disciplinary. It involves activities
of traffic scientists, psychologists, sociologists, biologists, physicists, computer scientists, and others.
Therefore, it is not surprising that there are sometimes different or even controversial views on the subject, e.g. with regard to the concept of ``panic'', the explanation of collective, spatio-temporal patterns of motion in pedestrian crowds, the best modeling concept, or the optimal number of parameters of a model.
\par
In this contribution, we will start with a short history of pedestrian modeling and, then, 
introduce the wide-spread ``social force model'' of pedestrian interactions
to illustrate further issues such as, for example, model calibration  by video tracking data.
Next, we will turn to the subject of crowd dynamics, since one typically finds the formation 
of large-scale spatio-temporal patterns of motion, when many pedestrians interact with each other. 
These patterns will be discussed in some detail
before we will turn to evacuation situations and cases of extreme densities, where one can sometimes observe 
the breakdown of coordination. Finally, we will address possibilities to design improved pedestrian 
facilities, using special evolutionary algorithms. 

\section{Pedestrian Dynamics}

\subsection{Short History of Pedestrian Modeling}

Pedestrians have been empirically
studied for more than four decades \cite{hist6,Old68,eng2}.
The evaluation methods initially applied were based on direct observation,
photographs, and time-lapse films. 
For a long time, the main goal of these studies was to develop 
a {\em level-of-service concept} \cite{LOS0}, 
{\em design elements} of pedestrian facilities
\cite{design3,Whyte,Hel97a,TranSci}, or {\em planning guidelines} \cite{eng1,Tra85}. 
The latter have usually the form of {\em regression relations}, which are, however, not very
well suited for the prediction of pedestrian flows in pedestrian zones
and buildings with an exceptional architecture, or in challenging
evacuation situations. Therefore, a number of
simulation models have been proposed, e.g. {\em queueing models} 
\cite{queue3}, {\em transition matrix models} \cite{Garbrecht},
and {\em stochastic models} \cite{AshOleMcg76},
which are partly related to each other. In addition, there are
models for the {\em route choice behavior} of pedestrians
\cite{Timmer1,Hel92a}.
\par
None of these concepts adequately takes into account the
self-organization effects occuring in pedestrian crowds.
These are the subject of recent experimental studies 
\cite{TranSci,Lattice,Daamen,Isobe,Seyfried,Kretz2}.
Most pedestrian models, however, were formulated before.
A first modeling approach that appears to be suited to reproduce spatio-temporal
patterns of motion was proposed by Henderson \cite{fluid}, 
who conjectured that pedestrian crowds behave 
similar to gases or fluids (see also \cite{Hug01}). 
This could be partially confirmed, but a realistic gas-kinetic or fluid-dynamic
theory for pedestrians must contain corrections due to their 
particular interactions (i.e. avoidance and deceleration maneuvers) which, of course, do not
obey momentum and energy conservation. Although such a theory can be actually
formulated \cite{Hel92b,Hoog2Bo00a}, for practical applications 
a direct simulation of {\em individual} pedestrian motion is favourable,
since this is more flexible. As a consequence, pedestrian research mainly focusses on 
{\em agent-based models} of pedestrian crowds, which also allow one to consider local 
coordination problems. The ``social force model'' \cite{Hel91,Pedestrians} is maybe the most 
well-known of these models, but we also like to mention {\em cellular automata} of pedestrian dynamics 
\cite{GipMa85,Bolay,BlueAd98,FukIs99a,MurIrNa99,KlMeyeWaSc00,BurKlaSchZi01} and
{\em AI-based models} \cite{Wayfinding,Reynolds1}. 

\subsection{The Social Force Concept}

In the following, we shall shortly introduce the social force concept, which reproduces 
most empirical observations in a simple and natural way.
Human behavior often seems to be ``chaotic'', irregular, and unpredictable. So, 
why and under what conditions can we model it by means of forces?
First of all, we need to be confronted with a phenomenon of motion in some
(quasi-)continuous space, which may be also an abstract behavioral space such as an opinion
scale \cite{behsci}. Moreover, it is favourable to have a system where the fluctuations due to
unknown influences are not large compared to the systematic,
deterministic part of motion. This is usually the case in pedestrian 
traffic, where people are confronted with standard situations and react
``automatically'' rather than taking complicated decisions,  e.g. if they have to evade others.  
\par
This ``automatic'' behavior can be interpreted as the result of a
{\em learning process} based on trial and error
\cite{HelMolFaBo00}, which can be simulated with {\em evolutionary algorithms} 
\cite{Klock}.  For example, pedestrians have a preferred side of walking, since an 
asymmetrical avoidance behavior turns out to be profitable \cite{Hel91,HelMolFaBo00}. The related
{\em formation of a behavioral convention} can be described by means 
of {\em evolutionary game theory} \cite{Hel91,Hel92c}.
\par
Another requirement is the vectorial additivity of the separate force terms
reflecting different environmental influences. This is probably an approximation, but
there is some experimental evidence for it. Based on quantitative measurements for animals and
test persons subject to separately or simultaneously applied
stimuli of different nature and strength, one could show that the behavior in
conflict situations can be described by a superposition of forces \cite{Mil44,Mil59}.
This fits well into a concept by Lewin \cite{Lewin}, according to which behavioral changes
are guided by so-called {\em social fields} or {\em social forces}, which has later on 
been put into mathematical terms 
\cite{Hel91,Hel94}. In some cases, social
forces, which determine the amount and direction 
of systematic behavioral changes, can be expressed as gradients of dynamically
varying potentials, which reflect the social or behavioral fields resulting from
the interactions of individuals. 
Such a social force concept was applied to
opinion formation and migration \cite{Hel94}, 
and it was particularly successful in the description of collective pedestrian 
behavior \cite{TranSci,Hel91,Pedestrians,HelMolFaBo00}.
\par
For reliable simulations of pedestrian crowds, we do not
need to know whether a certain pedestrian, say, turns to the right at 
the next intersection. It is sufficient to have a good estimate 
what percentage of pedestrians turns to the right. This can be
either empirically measured or estimated by means of route choice 
models \cite{Timmer1}. 
In some sense, the uncertainty about the individual behaviors is
averaged out at the macroscopic level of description. 
Nevertheless, we will use the more flexible microscopic
simulation approach based on the social force concept. According to this,
the temporal change of the location $\vec{r}_\alpha(t)$ of pedestrian $\alpha$ obeys the
equation
\begin{equation}
 \frac{d\vec{r}_\alpha(t)}{dt} = \vec{v}_\alpha(t) \, .
\label{motion}
\end{equation}
Moreover, if $\vec{f}_\alpha(t)$ denotes the sum of social forces influencing pedestrian $\alpha$ and if 
$\vec{\xi}_\alpha(t)$ are individual fluctuations reflecting unsystematic behavioral variations,
the velocity changes are given by the {\em acceleration equation}
\begin{equation}
  \frac{d\vec{v}_\alpha}{dt} = \vec{f}_\alpha(t) + \vec{\xi}_\alpha(t) \, .
 \label{acceleration}
\end{equation}
A particular advantage of this approach is that we can take into account
the flexible usage of space by pedestrians, requiring a continuous treatment of motion.
It turns out that this point is essential to reproduce the empirical observations in a natural 
and robust way, i.e. without having to adjust the model to each single situation and measurement site.
Furthermore, it is interesting to note that, if the fluctuation term is neglected, 
the social force model can be interpreted as a particular {\it differential
game}, i.e. its dynamics can be derived from the minimization of a special utility function \cite{Hoogendoorn}.

\subsection{Specification of the Social Force Model}\label{Model}

The social force model for pedestrians assumes that 
each individual $\alpha$ is trying to move in a desired direction $\vec{e}_{\alpha}^0$ with a 
desired speed $v_{\alpha}^0$, and that it adapts the actual velocity $\vec{v}_{\alpha}$ to the
desired one, $\vec{v}_\alpha^0 = v_\alpha^0 \vec{e}_\alpha^0$, within 
a certain relaxation time $\tau_{\alpha}$. 
The systematic part $\vec{f}_\alpha(t)$ of the acceleration force of pedestrian $\alpha$ is then given by
\begin{equation}
\vec{f}_{\alpha}(t)=\frac{1}{\tau_{\alpha}}(v_{\alpha}^0\vec{e}_{\alpha}^0-\vec{v}_{\alpha})
+\sum_{\beta(\ne\alpha)}\vec{f}_{\alpha\beta}(t)+\sum_{i}\vec{f}_{\alpha i}(t) \, ,
\end{equation}
where the terms $\vec{f}_{\alpha\beta}(t)$ and $\vec{f}_{\alpha i}(t)$ denote the repulsive forces
describing attempts to keep a certain safety distance to other pedestrians $\beta$ and obstacles $i$.
In very crowded situations, additional physical contact forces come into play (see Sec. \ref{Gran1}). Further forces
may be added to reflect attraction effects between members of a group or other influences. For details
see Ref. \cite{HelMolFaBo00}.
\par
First, we will assume a simplified interaction force of the form
\begin{equation}
 \vec{f}_{\alpha\beta}(t) =  \vec{f}\big(\vec{d}_{\alpha\beta}(t)\big) \, ,
\end{equation}
where $\vec{d}_{\alpha\beta} = \vec{r}_\alpha - \vec{r}_\beta$ is the distance vector pointing from 
pedestrian $\beta$ to $\alpha$. Angular-dependent shielding effects may be furthermore taken into account by a 
prefactor describing the anisotropic reaction to situations in front of as compared to behind a pedestrian
\cite{Pedestrians,calibration}, see Sec. \ref{ang}. However, we will start with a {\bf circular specification}
of the distance-dependent interaction force, 
\begin{equation}
 \vec{f}(\vec{d}_{\alpha\beta}) = A_\alpha \mbox{e}^{- d_{\alpha\beta}/B_\alpha} 
\frac{\vec{d}_{\alpha\beta}}{\|\vec{d}_{\alpha\beta}\|} \, ,
\label{g}
\end{equation}
where $d_{\alpha\beta} = \|\vec{d}_{\alpha\beta}\|$ is the distance. The parameter  
$A_\alpha$ reflects the {\em interaction 
strength}, and $B_\alpha$ corresponds to the {\em interaction range}. While the dependence
on $\alpha$ explicitly allows for a dependence of these parameters on the single individual,
we will assume a homogeneous population, i.e.
$A_\alpha = A$ and $B_\alpha = B$ in the following. Otherwise, it would be hard
to collect enough data for parameter calibration.
\par
{\bf Elliptical specification:} Note that it is possible to express Eq. (\ref{g}) as gradient of
an exponentially decaying potential $V_{\alpha \beta}$. This circumstance can be used to formulate 
a generalized, elliptical interaction force via the potential
\begin{equation}
V_{\alpha\beta}(b_{\alpha\beta})=AB \, \mbox{e}^{-b_{\alpha\beta}/B} \, ,
\end{equation}
where the variable $b_{\alpha\beta}$ denotes the semi-minor axis $b_{\alpha\beta}$ of the elliptical
equipotential lines.  This has been specified according to
\begin{equation}
2b_{\alpha\beta} =\sqrt{(\|\vec{d}_{\alpha \beta}\| + \|\vec{d}_{\alpha\beta}-(\vec{v}_\beta - \vec{v}_\alpha)
\Delta t \|)^2 - \|(\vec{v}_\beta - \vec{v}_\alpha)\Delta t\|^2} \, ,
\label{two}
\end{equation}
so that both pedestrians $\alpha$ and $\beta$ are treated symmetrically.
The repulsive force is related to the above potential via
\begin{equation}
\vec{f}_{\alpha \beta}(\vec{d}_{\alpha \beta})=-\vec{\nabla}_{\vec{d}_{\alpha \beta}} V_{\alpha \beta}(b_{\alpha\beta}) 
 = - \frac{dV_{\alpha \beta}(b_{\alpha\beta})}{db_{\alpha\beta}} \vec{\nabla}_{\vec{d}_{\alpha \beta}} b_{\alpha\beta}(\vec{d}_{\alpha\beta})\, ,
\end{equation}
where $\vec{\nabla}_{\vec{d}_{\alpha \beta}}$ represents the gradient with respect to $\vec{d}_{\alpha \beta}$.
Considering the chain rule, $\|\vec{z}\| = \sqrt{\vec{z}^2}$, and $\vec{\nabla}_{\vec{z}}
\|\vec{z}\| = \vec{z}/\sqrt{\vec{z}^2} = \vec{z}/\|\vec{z}\|$, this leads to the explicit formula
\begin{equation}
 \vec{f}_{\alpha \beta}(\vec{d}_{\alpha \beta}) = A \mbox{e}^{-b_{\alpha\beta}/B} \cdot
 \frac{\|\vec{d}_{\alpha\beta}\| + \|\vec{d}_{\alpha\beta}-\vec{y}_{\alpha\beta}\| }{2b_{\alpha\beta}}
\cdot \frac{1}{2} \left( \frac{\vec{d}_{\alpha\beta}}{\|\vec{d}_{\alpha\beta}\|} 
 + \frac{\vec{d}_{\alpha\beta}-\vec{y}_{\alpha\beta}}{\|\vec{d}_{\alpha\beta}-\vec{y}_{\alpha\beta}\|}\right)  
\end{equation}
with $\vec{y}_{\alpha\beta} = (\vec{v}_\beta - \vec{v}_\alpha) \Delta t$. We used $\Delta t = 0.5$s.
For $\Delta t = 0$, we regain the expression of Eq.~(\ref{g}). 
\par
The elliptical specification has two major advantages 
compared to the circular one: First, the interactions depend not only on the distance, but also on the relative velocity.
Second, the repulsive force is not strictly directed from pedestrian $\beta$ to pedestrian $\alpha$,
but has a lateral component. As a consequence, this leads to less confrontative, smoother (``sliding'') evading maneuvers.
Note that further velocity-dependent
specifications of pedestrian interaction forces have been proposed \cite{Hel97a,Pedestrians}, but we will restrict to the
above specifications, as these are sufficient to demonstrate the method of evolutionary model calibration. For suggested improvements regarding the specification of social forces see, for example, 
Refs. \cite{Taras,Lippert}.

\subsection{Angular Dependence}\label{ang}

In reality, of course, pedestrian interactions are not isotropic, but dependent on the
angle $\varphi_{\alpha\beta}$ of the encounter, which is given by the formula
\begin{equation}
 \cos(\varphi_{\alpha\beta}) = \frac{\vec{v}_\alpha}{\|\vec{v}_\alpha\|} 
 \cdot \frac{-\vec{d}_{\alpha\beta}}{\|\vec{d}_{\alpha\beta}\|} \, .
\end{equation}
Generally, pedestrians show little response to pedestrians behind them. This
can be reflected by an angular-dependent prefactor $w(\varphi_{\alpha\beta})$ of the 
interaction force  \cite{calibration}.
Empirical results are represented in Fig.~\ref{expo} (right). Reasonable results are
obtained for the following specification of the prefactor:
\begin{equation}
w\big(\varphi_{\alpha\beta}(t)\big) 
=\left(\lambda_{\alpha}+(1-\lambda_{\alpha})\frac{1+\cos(\varphi_{\alpha\beta})}{2}\right) \, ,
\label{w}
\end{equation}
where $\lambda_\alpha$ with $0 \le \lambda_\alpha \le 1$
is a parameter which grows with the strength of interactions from behind. An evolutionary
parameter optimization gives values $\lambda \approx 0.1$ (see Sec. \ref{EvCal}, i.e. a strong anisotropy. 
Other angular-dependent specifications split up the interaction
force between pedestrians into a component against the direction of motion and another one perpendicular to it. Such a description allows for even smoother avoidance maneuvres. 

\subsection{Evolutionary Calibration with Video Tracking Data}\label{EvCal}

\begin{figure}[htb] \begin{center}
    \includegraphics[width=0.4\textwidth]{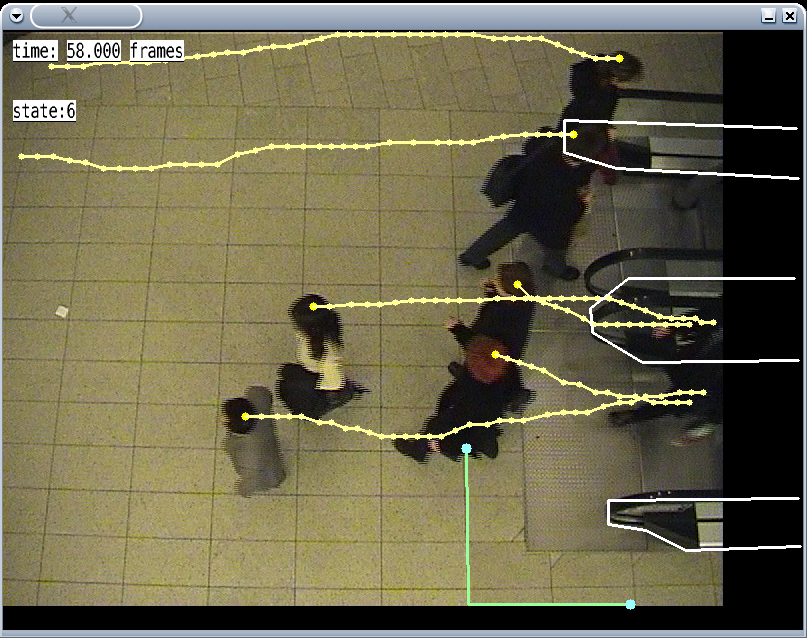}
    \includegraphics[width=0.45\textwidth]{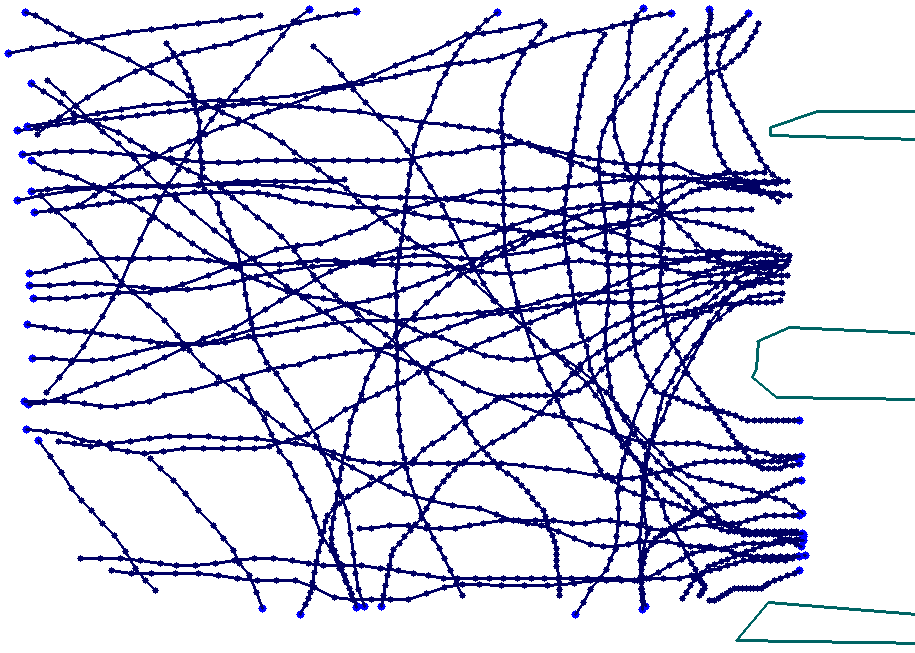}
    \caption{Video tracking used to extract the trajectories of pedestrians from video recordings close to
two escalators (after \cite{calibration}). 
Left: Illustration of the tracking of pedestrian heads. 
Right: Resulting trajectories after being transformed onto the two-dimensional plane.}  
\label{Fig2}
\end{center} 
\end{figure}

For parameter calibration, several video recordings of pedestrian crowds in different natural environments
have been used. The dimensions of
the recorded areas were known, and the floor tiling or environment provided something like a
``coordinate system''. The heads were automatically determined by seaching for round moving structures,
and the accuracy of tracking was improved by comparing actual with linearly extrapolated positions
(so it would not happen so easily that the algorithm interchanged or ``lost'' closeby pedestrians). 
The trajectories of the heads were then projected on two-dimensional space in a way correcting for
distortion by the camera perspective. A representative plot of the resulting trajectories 
is shown in Fig.~\ref{Fig2}. Note that trajectory data have been obtained with infra-red sensors \cite{KerridgeEmpirical} 
or video cameras \cite{Hoog,Teknomo} for several years now, but algorithms that can simultaneously handle
more than one thousand pedestrians have become available only recently \cite{turb}. 
\par
For model calibration, it is recommended to use a hybrid method fusing empirical trajectory data
and microscopic simulation data of pedestrian movement in space. In corresponding algorithms,
a virtual pedestrian is assigned to each tracked pedestrian in the simulation domain. 
One then starts a simulation for a time period $T$ (e.g. 1.5 seconds), in which one pedestrian 
$\alpha$ is moved according to a simulation of the social force model, while the others are 
moved exactly according to the trajectories extracted from the videos.
This procedure is performed for all pedestrians $\alpha$
and for several different starting times $t$, using a fixed parameter set for the social force model.
\par
Each simulation run is performed according to the following scheme:
\begin{enumerate}
	\item Define a starting point and calculate the state 
(position $\vec{r}_{\alpha}$, velocity $\vec{v}_{\alpha}$, and acceleration 
$\vec{a}_{\alpha} = d\vec{v}_\alpha/dt$) for each pedestrian $\alpha$.
	\item Assign a desired speed $v_\alpha^0$ to each pedestrian, e.g. 
the maximum speed during the 
pedestrian tracking time. This is sufficiently accurate, if the overall pedestrian density is not
too high and the desired speed is constant in time. 
	\item Assign a desired goal point for each pedestrian, e.g. the end point of the trajectory.
	\item Given the tracked motion of the surrounding pedestrians $\beta$,
simulate the trajectory of pedestrian $\alpha$ over a time period $T$ based on 
the social force model, starting at the actual location $\vec{r}_\alpha(t)$.
\end{enumerate}
After each simulation run, one determines the relative distance error
\begin{equation}
\frac{\|\vec{r}_\alpha^{\rm simulated}(t+T) - \vec{r}_\alpha^{\rm tracked}(t+T)\|}
 {\|\vec{r}_\alpha^{\rm tracked}(t+T) - \vec{r}_\alpha^{\rm tracked}(t)\|} \, .
\end{equation}
After averaging the relative distance errors over the pedestrians $\alpha$ and starting times $t$, 
1 minus the result can be taken as measure of the goodness of fit (the ``fitness'') 
of the parameter set used in the pedestrian simulation.
Hence, the best possible value
of the ``fitness'' is 1, but any deviation from the real pedestrian trajectories implies lower values.
\par\begin{table}[htbp]
\begin{center}
\begin{tabular}{|l||c|c|c||r|}
\hline
Model & A [m/s$^2$] & B [m] & $\lambda$ & Fitness \\
\hline
Extrapolation & 0 & -- &  -- & 0.34 \\
Circular & 0.42 $\pm$ 0.26 & 1.65 $\pm$ 1.01 & 0.12 $\pm$ 0.07 & 0.40 \\
Elliptical & 0.04 $\pm$ 0.01 & 3.22 $\pm$ 0.67 & 0.06 $\pm$ 0.04 & 0.61 \\
\hline
\end{tabular}
\end{center}
\caption[]{Interaction strength $A$ and interaction range $B$ 
resulting from our evolutionary parameter calibration 
for the circular and elliptical specification of the interaction forces between pedestrians (see main text),
with an assumed angular dependence according to Eq. (\ref{w}).
A comparison with the extrapolation scenario, which assumes constant speeds, allows one to
judge the improvement in the goodness of fit (``fitness'') by the specified interaction force.
The calibration was based on three different video recordings, one for low crowd density, one for medium, and one
for high density (see Ref. \cite{calibration} for details). The parameter values are specified as mean 
value $\pm$ standard deviation. The best fitness value obtained with the elliptical specification
for the video with the lowest crowd density was as high as 0.9.}
\label{Tab}
\end{table}
One result of such a parameter optimization is that, for each video, 
there is a broad range of parameter combinations of $A$ and $B$ which perform almost 
equally well \cite{calibration}.  This allows one to apply additional
goal functions in the parameter optimization, e.g. to determine among the
best performing parameter values such parameter combinations, which perform well for 
{\em several} video recordings, using a fitness function which equally weights the fitness reached in
each single video. This is how the parameter values listed in Table~\ref{Tab} were determined. 
It turns out that, in order to reach a good model
performance, the pedestrian interaction force must be specified velocity dependent, as in the elliptical model.
\par
\begin{figure}[htbp] 
\begin{center}
    \includegraphics[height=4.6cm]{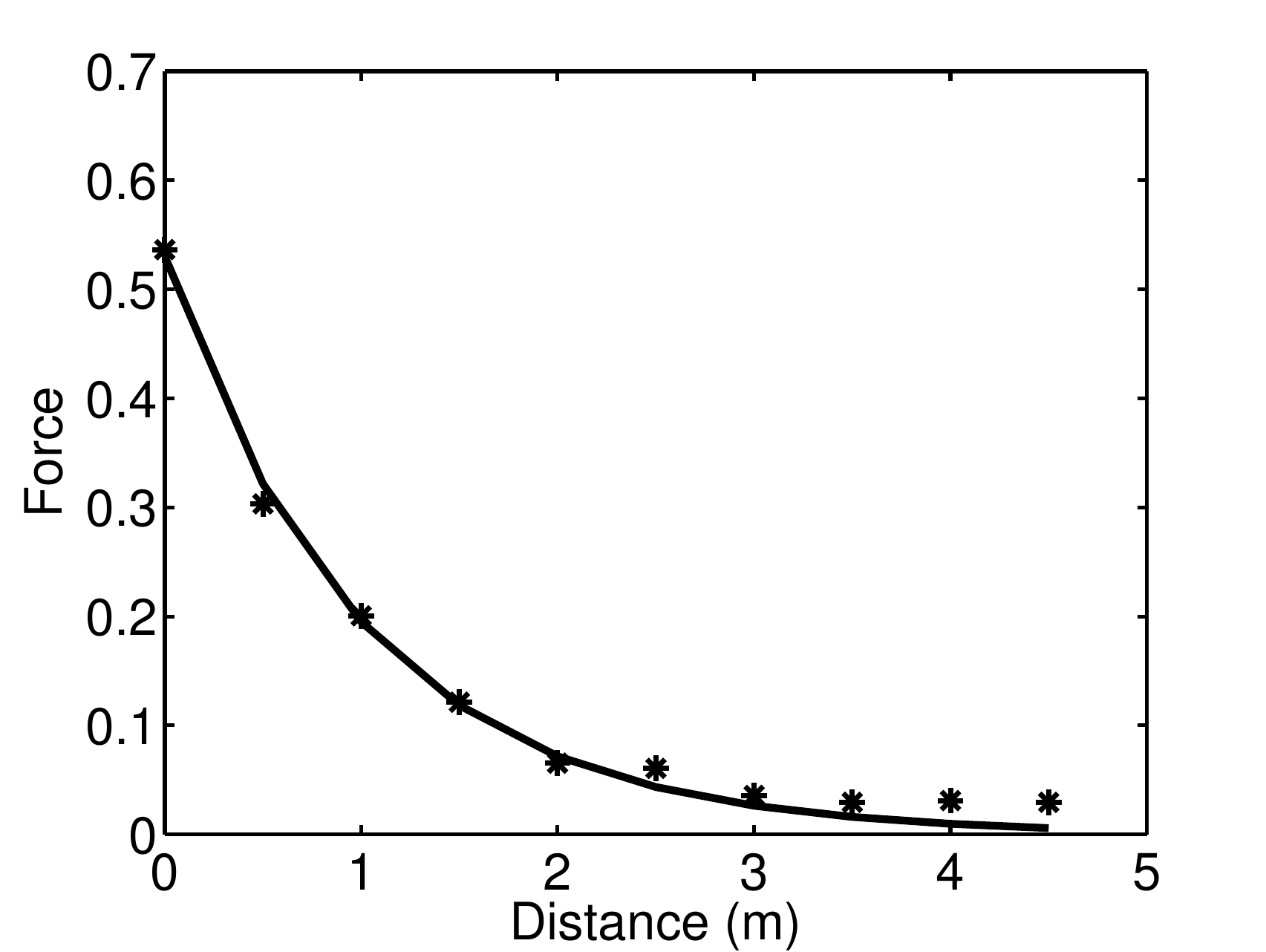}
    \includegraphics[height=4.4cm]{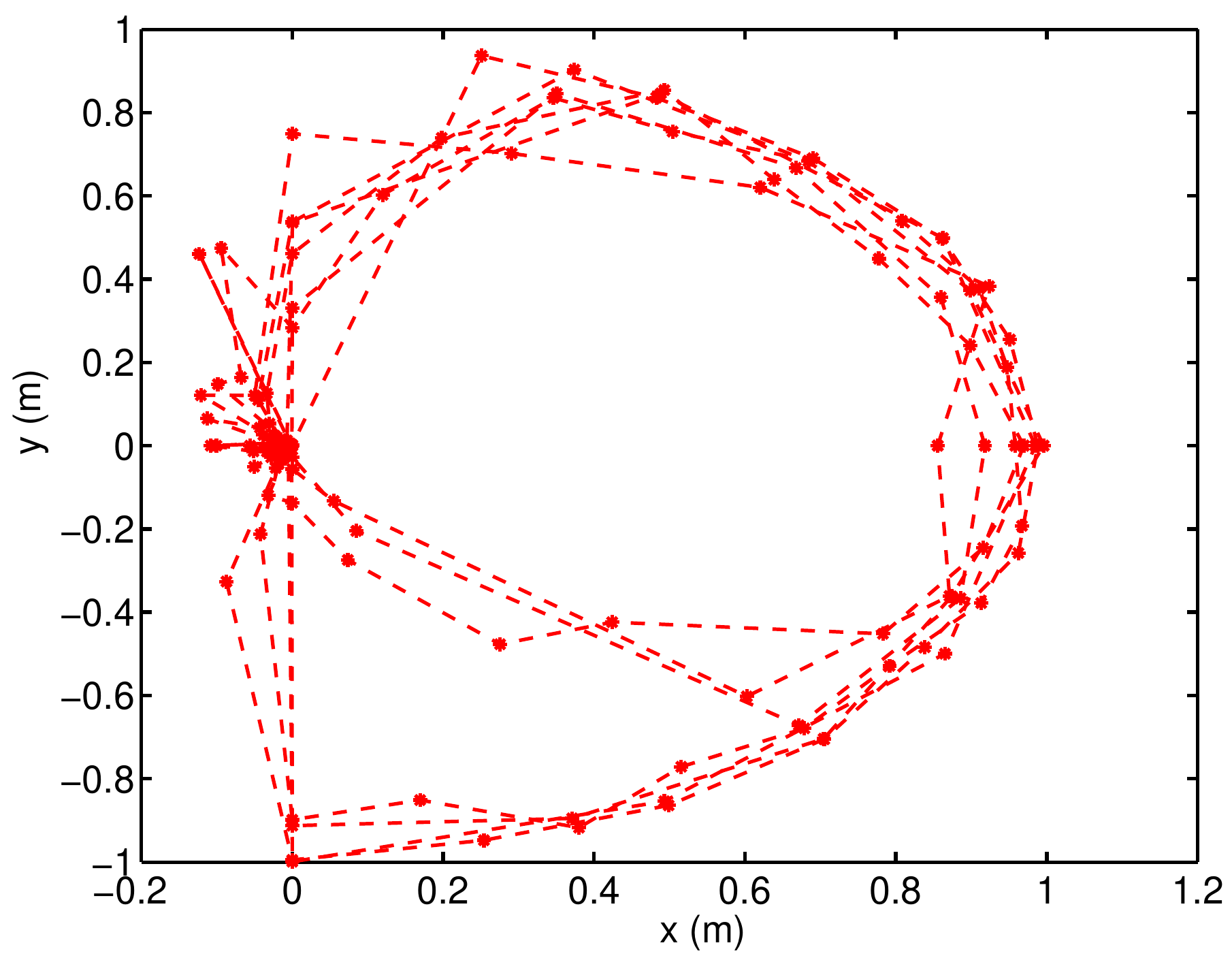}
\vspace*{-5mm}
\end{center}
\caption{Results of an evoluationary fitting of pedestrian interactions.
Left: Empirically determined distance dependence of the interaction force between pedestrians. An exponential decay fits the empirical data quite well. The dashed fit curve corresponds to 
Eq.~(\ref{g}) with the parameters $A=0.53$ and $B=1.0$. 
Right: Angular dependence of the influence of other
pedestrians. The direction along the positive $x$ axis corresponds to the
walking direction of pedestrians, $y$ to the perpendicular direction. (After \cite{calibration}.)}
\label{expo}
\end{figure}

Note that our evolutionary fitting method can be also used to determine interaction laws without
prespecified interaction functions. For example, 
one can obtain the distance dependence of pedestrian interactions without
a pre-specified function. For this, one adjusts the values of the force at given distances
$d_k = k d_1$ (with $k\in \{1,2,3,...\}$) in an evolutionary way. 
To get some smoothness, linear interpolation is applied. The resulting fit curve is presented 
in Fig.~\ref{expo} (left). It turns out that the empirical dependence of the force with distance can be 
well fitted by an exponential decay.
\par

\section{Crowd Dynamics}

\subsection{Analogies with gases, fluids, and granular media}\label{analogies}

When the density is low, pedestrians can move freely, and the observed crowd dynamics can be
partially compared with the behavior of gases. At medium and high densities, however,
the motion of pedestrian crowds shows some striking analogies
with the motion of fluids: 
\begin{enumerate}
\item Footprints of pedestrians in snow look similar to streamlines of fluids
\cite{Hel92a}. 
\item At borderlines between opposite directions of walking one can observe
``viscous fingering'' \cite{fingering,StanOs86}. 
\item The emergence of pedestrian streams through 
standing crowds \cite{Hel97a,HelMolFaBo00,Arns} appears analogous to the
formation of river beds \cite{Stol96,RodRi97}.
\end{enumerate}
At high densities, however, the observations have rather analogies with
driven granular flows. This will be elaborated in more detail in Secs.~\ref{Gran1} and \ref{Gran2}.
In summary, one could say that fluid-dynamic analogies work reasonably well in normal situations,
while granular aspects dominate at extreme densities. Nevertheless, the analogy is limited,
since the self-driven motion and the violation of momentum conservation
imply special properties of pedestrian flows. For example, one usually does not observe eddies.

\subsection{Self-Organization of Pedestrian Crowds}\label{selfo}

Despite its simplifications, the social force model of pedestrian dynamics
describes a lot of observed phenomena quite realistically. Especially, it allows one to
explain various self-organized spatio-temporal patterns that are not externally
planned, prescribed, or organized, e.g. by traffic signs, laws, or behavioral
conventions \cite{Hel97a,TranSci,HelMolFaBo00}. Instead, the spatio-temporal patterns discussed below emerge due to the non-linear interactions of pedestrians even without assuming strategical
considerations, communication, or imitative behavior of pedestrians. Despite this,
we may still interpret the forming cooperation patterns as phenomena that 
establish social order on short time scales. It is actually surprising that strangers coordinate 
each other within seconds, if they have grown up in a similar enviroment. People from different countries, however,
are sometimes irritated about local walking habits, which indicates that learning effects and cultural
backgrounds still play a role in social interactions as simple as random pedestrian encounters. Rather than on
particular features, however, in the following we will focus on the common, internationally reproducible observations.
\begin{figure}[!ht]
\unitlength1cm
\begin{center}
\begin{picture}(12,7.5)
\put(0,0){\includegraphics[width=6cm]{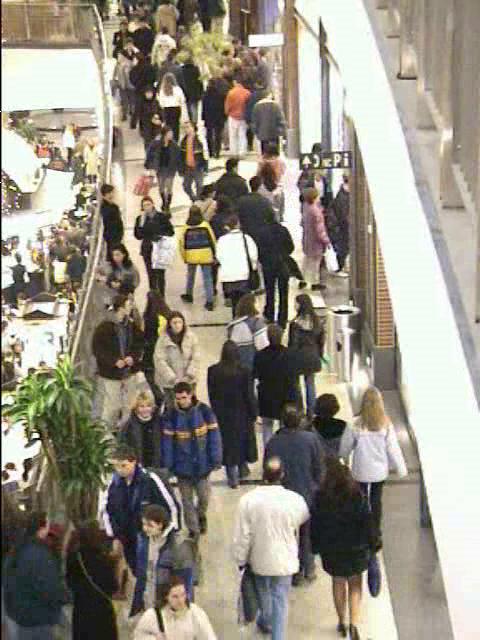}} 
\put(6.2,8.2){\includegraphics[height=6cm,angle=-90]{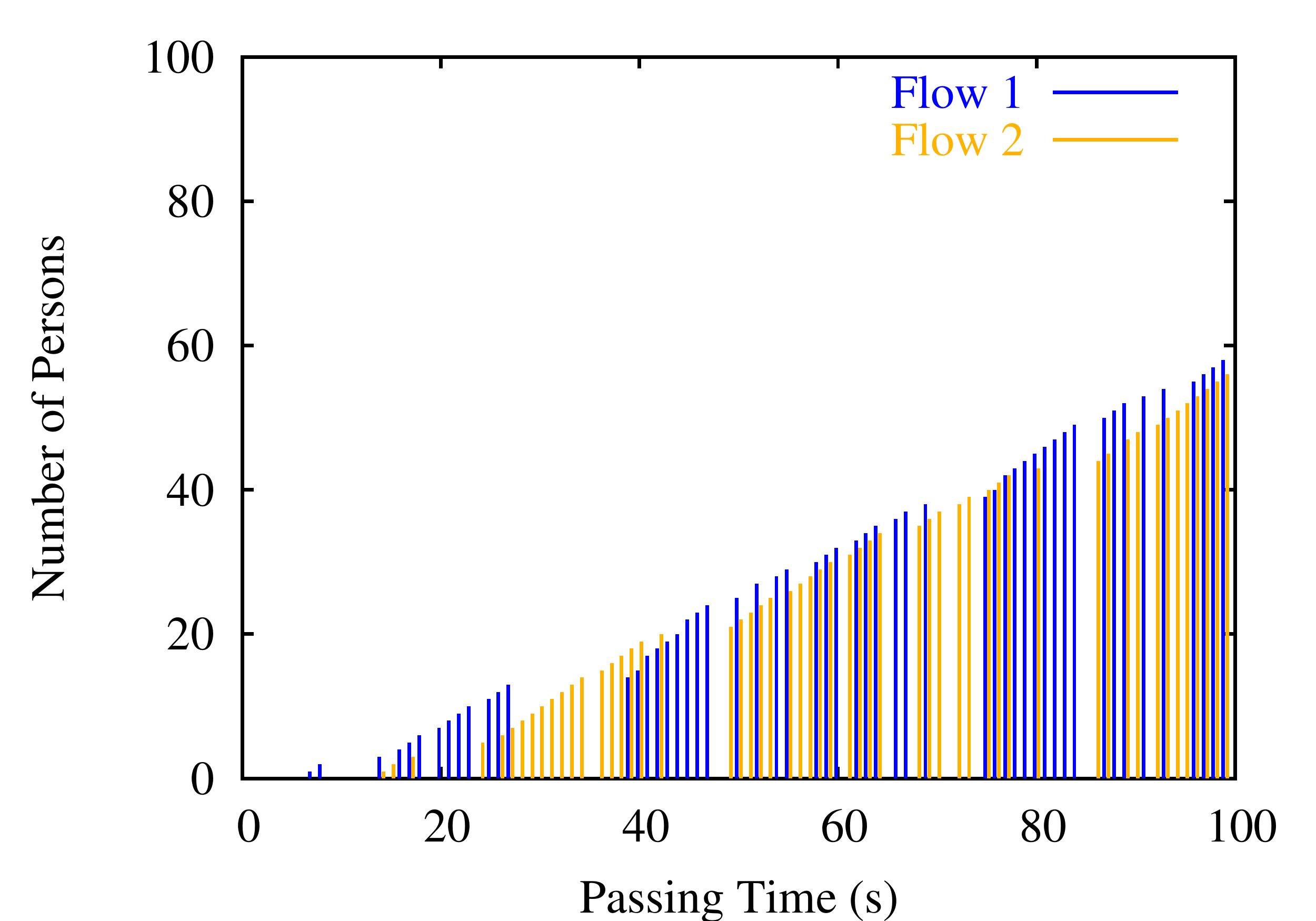}}
\put(6.2,-0.2){\includegraphics[width=6cm]{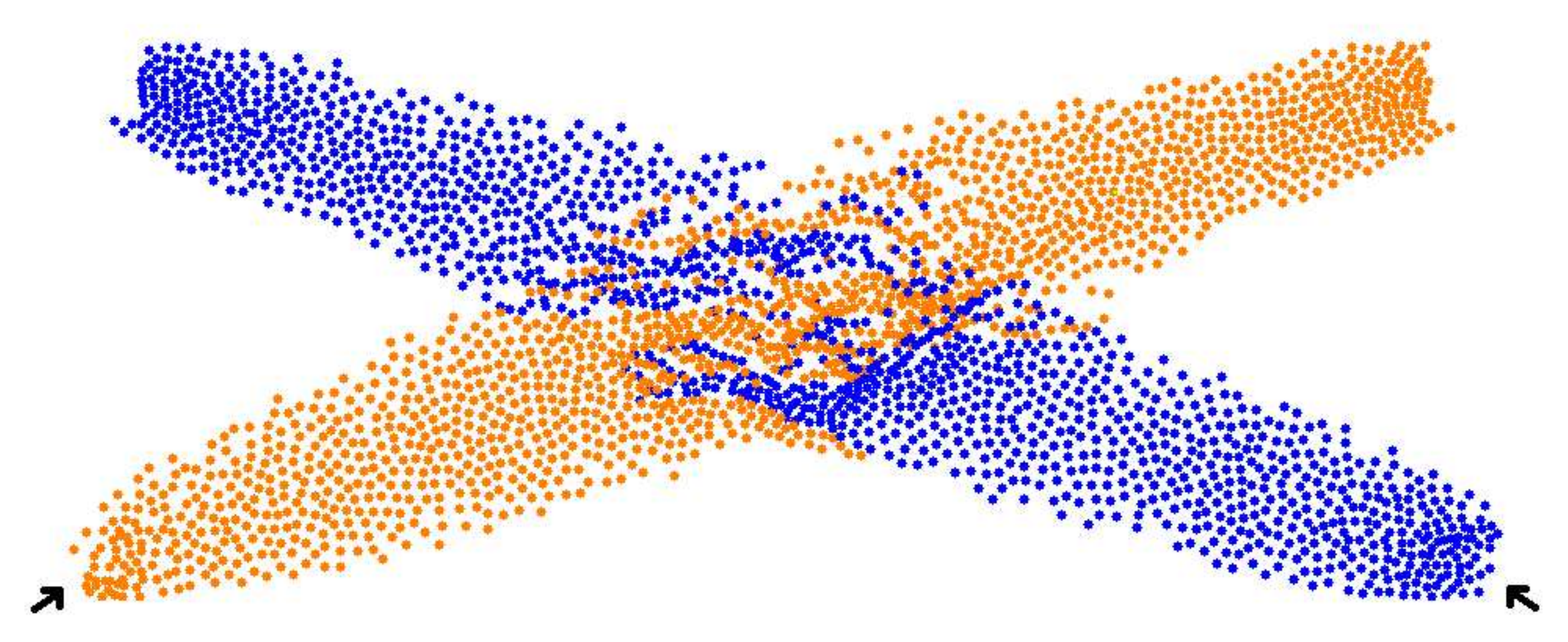}} 
\put(6.2,1.8){\includegraphics[width=6cm]{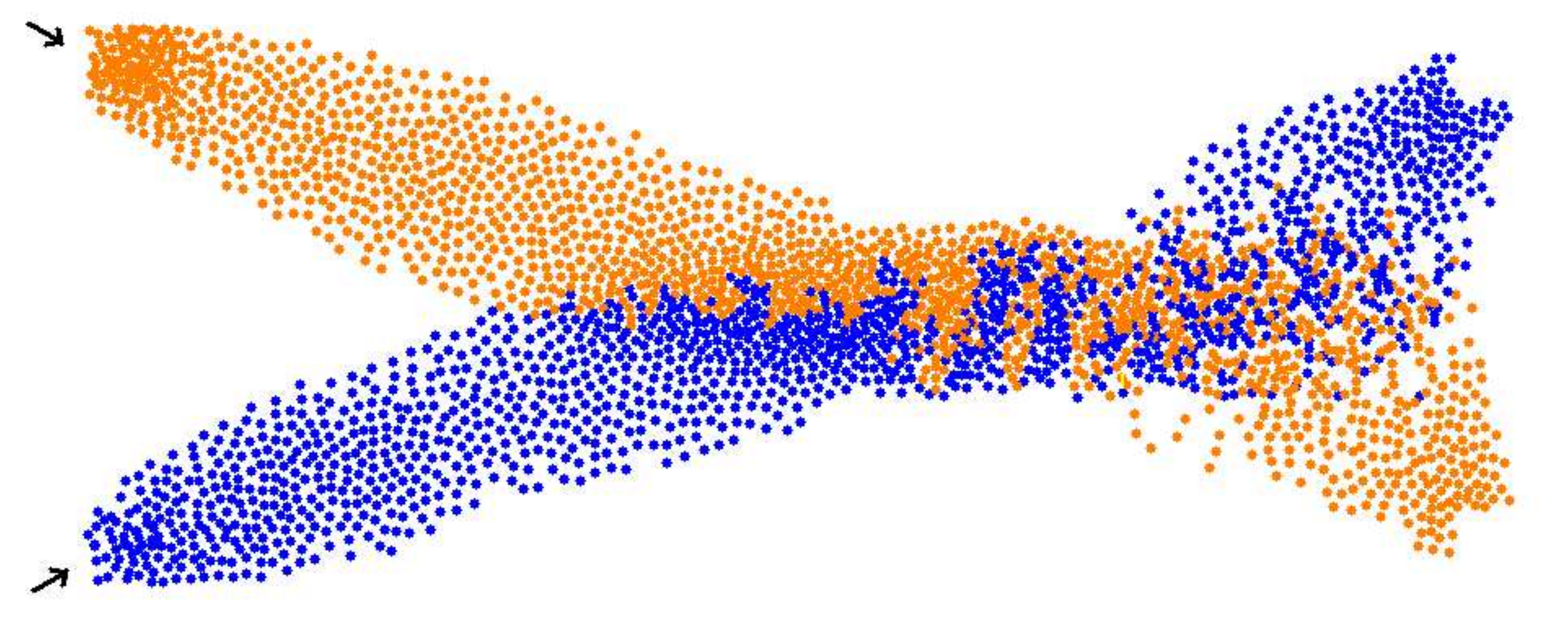}} 
\end{picture}
\end{center}
\caption[]{Self-organization of pedestrian crowds. Left: Photograph of lanes formed in a shopping center.
Computer simulations reproduce the self-organization of such lanes very well. Top right:
Evaluation of the cumulative number of pedestrians passing a bottleneck from different sides. 
One can clearly see that the narrowing is often passed by groups of people in an oscillatory way rather than one by one.
Bottom right: Multi-agent simulation of two crossing pedestrian streams, showing the phenomenon
of stripe formation. This self-organized pattern allows pedestrians to pass 
the other stream without having to stop, namely by moving sidewards in a forwardly moving
stripe.  (After Ref. \cite{TranSci}.)\label{patterns}}
\end{figure}

\subsubsection{Lane formation:}

In pedestrian flows one can often observe that oppositely moving pedestrians 
are forming lanes of uniform walking direction (see Fig.~\ref{patterns}) \cite{TranSci,Kretz2,Hel91,Pedestrians}. 
This phenomenon even occurs when there is not a large distance to separate each other,
e.g. on zebra crossings. However, the width of lanes increases (and their number decreases), if the interaction
continues over longer distances (and if perturbations, e.g. by flows entering or leaving on the sides, are low;
otherwise the phenomenon of lane formation may break down \cite{HelFaVi00a}).
\par
Lane formation may be viewed as {\it segregation phenomenon} \cite{Schelling,Platkowski}. Although there
is a weak preference for one side (with the corresponding behavioral convention depending on the country),
the observations can only be well reproduced when repulsive pedestrian interactions are taken into account.
The most relevant factor for the lane formation phenomenon is the higher relative velocity of
pedestrians walking in opposite directions. Compared to people following each other, oppositely
moving pedestrians have more frequent interactions until they have segregated into
separate lanes by stepping aside whenever another pedestrian is encountered. 
The most long-lived patterns of motion are the ones which change the least.
It is obvious that such patterns correspond to lanes, as they minimize the 
frequency and strength of avoidance maneuvers. Interestingly enough, as computer simulations show,
lane formation occurs also when there is no preference for any side.
\par
Lanes minimize frictional effects, accelerations, energy consumption, and delays 
in oppositely moving crowds. Therefore, one could say that they are
a pattern reflecting ``collective intelligence''. In fact, it is not possible for a single pedestrian to
reach such a collective pattern of motion. Lane formation is a self-organized collaborative pattern of motion
originating from simple pedestrian interactions. Particularly in cases of no side preference, 
the system behavior cannot be understood by adding up the behavior of the single individuals. This 
is a typical feature of complex, self-organizing systems and, in fact, a wide-spread characteristics of social 
systems. It is worth noting, however, that it does not require a conscious behavior to reach forms of social organization
like the segregation of oppositely moving pedestrians into lanes. This organization occurs automatically, 
although most people are not even aware of the existence of this phenomenon. 

\subsubsection{Oscillatory flows at bottlenecks:}

At bottlenecks, bidirectional flows of moderate density are often characterized by oscillatory changes
in the flow direction (see Fig.~\ref{patterns}) \cite{TranSci,Pedestrians}. For example,  one can
sometimes observe this at entrances of museums during crowded art exhibitions
or at entrances of staff canteens during lunch time.
While these oscillatory flows may be interpreted as an effect of friendly behavior (``you go first, please''),
computer simulations of the social force model indicate that the collective behavior may again be understood by simple pedestrian interactions. That is, oscillatory flows can even occur in the absence of communication, although it may be involved in reality. The interaction-based mechanism of
oscillatory flows suggests to interpret them as 
another self-organization phenomenon, which again reduces frictional effects
and delays. That is, oscillatory flows have features of ``collective intelligence''. 
\par
While this may be interpreted as result
of a learning effect in a large number of similar situations (a ``repeated game''), our simulations suggest an even
simpler, ``many-particle'' interpretation:
Once a pedestrian is able to pass the narrowing, pedestrians
with the same walking direction can easily follow.
Hence, the number and ``pressure''
of waiting, ``pushy'' pedestrians on one side of the bottleneck
becomes less than on the other side. This eventually increases their
chance to occupy the passage. Finally,
the ``pressure difference'' is large enough to stop the flow and
turn the passing direction at the bottleneck. This reverses the situation, and
eventually the flow direction changes again, giving rise to oscillatory flows.
\par
At bottlenecks, further interesting observations can be made:
Hoogendoorn and Daamen \cite{HoDaa} report the formation of layers in unidirectional
bottleneck flows. Due to the partial overlap of neighboring layers, there is a zipper
effect. Moreover, Kretz {\em et al.} \cite{Kretz3} have observed that the specific flow through a narrow
bottleneck decreases with a growing width of the bottleneck, as long as it can be passed by one person at a time only. This is due to mutual obstructions, if two people are trying to enter the bottleneck simultaneously. If the opening is large enough to be entered by several people in parallel,
the specific flow stays constant with increasing width. Space is then used in a flexible way.

\subsubsection{Stripe formation in intersecting flows:}

In intersection areas, the flow of people often appears to be irregular or ``chaotic''. In fact, it can
be shown that there are several possible collective patterns of motion, among them rotary and
oscillating flows. However, these patterns continuously 
compete with each other, and a temporarily dominating pattern is destroyed by another one after a short
time. Obviously, there has not evolved any social convention that would establish and stabilize
an ordered and efficient flow at intersections. 
\par
Self-organized patterns of motion, however, are found in situations where pedestrian flows
cross each other only in two directions. In such situations, the phenomenon of stripe formation
is observed \cite{Ando}. Stripe formation 
allows two flows to penetrate each other without requiring the pedestrians to stop. 
For an illustration see Fig.~\ref{patterns}.
Like lanes, stripes are a segregation phenomenon, but not a stationary one. Instead, the stripes are
density waves moving into the direction of the sum of the directional vectors of both intersecting
flows. Naturally, the stripes extend sidewards into the direction which is 
perpendicular to their direction of motion. Therefore, the pedestrians move forward {\em with} the stripes and
sidewards {\em within} the stripes. 
Lane formation corresponds to the particular case of stripe formation 
where both directions are exactly opposite. In this case, no intersection
takes place, and the stripes do not move systematically. As in lane formation, stripe formation 
allows to minimize obstructing interactions and to maximize the average pedestrian speeds,
i.e. simple, repulsive pedestrian interactions again lead to an ``intelligent'' collective behavior.

\section{Evacuation Dynamics}

While the previous section has focussed on the dynamics of pedestrian crowds in normal
situations, we will now turn to the description of situations in which extreme crowd densities
occur. Such situations may arise at mass events, particularly in cases of urgent egress. While
most evacuations run relatively smoothly and orderly, the situation may also get out of control
and end up in terrible crowd disasters (see Tab. \ref{Tabl}). In such situations, one often speaks of ``panic'',
although, from a scientific standpoint, the use of this term is rather controversial. Here, however, we
will not be interested in the question whether ``panic'' actually occurs or not. We will rather focus on 
the issue of crowd dynamics at high densities and under psychological stress.

\subsection{Evacuation and Panic Research}

Computer models have been also developed for emergency and evacuation situations 
\cite{KlMeyeWaSc00,EvacSim,Ebihara,KetcCoWeMa93,OkaMa93,FireSim4,Stil00,ThomMa93,%
Lov98,Ham}. Most research into panic, however, has been of empirical nature 
(see, e.g. Refs.~\cite{Fire2,Hills,Canter}), carried out by social psychologists and others.
\par
With some exceptions,
panic is thought to occur in cases of scarce or dwindling resources \cite{Exp,Fire1},
which are either required for survival or anxiously desired. 
They are usually distinguished into escape panic (``stampedes'',
bank or stock market panic) and acquisitive panic
(``crazes'', speculative manias) \cite{Coll1,Coll3},
but in some cases this classification is questionable \cite{Stadion}.
\par
It is often stated that
panicking people are obsessed by short-term personal interests
uncontrolled by social and cultural constraints \cite{Fire1,Coll1}. This
is possibly a result of the reduced attention in situations of fear \cite{Fire1},
which also causes that options like side exits are mostly
ignored \cite{Fire2}. It is, however, mostly attributed to
social contagion \cite{Hills,Exp,Fire1,Coll1,Coll3,Stadion,SocPsy,SocPsy1,SocPsy2,Coll2,SocPsy4},
i.e., a transition from individual to mass psychology,
in which individuals transfer control over their actions to
others \cite{Coll3}, leading to conformity \cite{Bryan}. 
This ``herding behavior'' is in some sense irrational, as it often leads to bad
overall results like dangerous overcrowding and slower
escape \cite{Fire2,Coll3,Stadion}.
In this way, herding behavior can increase the
fatalities or, more generally, the damage in the crisis faced.
\par
The various socio-psychological theories for this contagion assume
hypnotic effects, rapport, mutual excitation of a primordial instinct, 
circular reactions, social facilitation (see the summary by Brown \cite{Coll2}), 
or the emergence of normative support for selfish behavior \cite{SocPsy4}.
Brown \cite{Coll2} and Coleman \cite{Coll3}
add another explanation related to the prisoner's dilemma \cite{Ax1,Ax2}
or common goods dilemma \cite{SciAm}, showing
that it is reasonable to make one's subsequent actions contingent upon
those of others. However, the socially favourable behavior of walking orderly is unstable,
which normally gives rise to rushing by everyone.
These thought\-ful considerations are well compatible
with many aspects discussed above and with the
classical experiments by Mintz \cite{Exp}, which showed
that jamming in escape situations depends on the reward structure
(``payoff matrix'').
\par
Nevertheless and despite of the frequent reports in the media and
many published investigations of crowd disasters (see Table~\ref{Tabl}), 
a quantitative understanding of the observed phenomena in panic stampedes was lacking
for a long time. The following sections will close this gap.
\begin{table}[htbp]
\caption[]{Incomplete list of major crowd disasters since 1970 after J. F. Dickie in Ref.~\cite{calibrate},
{\tt http://www.crowddynamics.com/Main/Crowddisasters.html},
{\tt http://SportsIllustrated.CNN.com/soccer/world/news/2000/07/09/stadium\_
disasters\_ap/}, and other internet sources, excluding fires, bomb attacks, 
and train or plane accidents.
The number of injured people was usually a multiple of the fatalities.\label{Tabl}}
\begin{center}{\footnotesize
\hspace*{-1mm}\begin{tabular}{|r|p{3.3cm}|p{2cm}|r|p{5.5cm}|}
\hline
Date  & Place  &                    Venue  &         Deaths & Reason\\
\hline
1971  & Ibrox, UK  &                 Stadium &             66 & Collapse of barriers \\
1974  & Cairo, Egypt &               Stadium &             48 & Crowds break barriers \\
1982  & Moscow, USSR  &              Stadium &            340 & Re-entering fans after last minute goal \\
1988 & Katmandu, Nepal & Stadium & 93 & Stampede due to hailstorm \\
1989  & Hillsborough, Sheffield, UK &  Stadium &  96 & Fans trying to force their way into the stadium \\
1990 & New York City & Bronx & 87 & Illegal happy land social club \\ 
1990  & Mena, Saudi Arabia &   Pedestrian Tunnel & 1426 & Overcrowding \\
1994 & Mena, Saudi Arabia &  Jamarat Bridge & 266 & Overcrowding \\ 
1996 & Guatemala~City,\hfill\, Guatemala & Stadium & 83 & Fans trying to force their way into the stadium \\
1998 & Mena, Saudi Arabia &  & 118 & Overcrowding \\
1999 & Kerala, India & Hindu Shrine & 51 & Collapse of parts of the shrine \\
1999 & Minsk, Belarus & Subway Station & 53 & Heavy rain at rock concert \\
2001 & Ghana, West Africa & Stadium & $>100$ & Panic triggered by tear gas\\
2004 & Mena, Saudi Arabia & Jamarat Bridge & 251 & Overcrowding\\
2005 & Wai, India & Religious Procession & 150 & Overcrowding (and fire)\\
2005 & Bagdad, Iraque & Religious Procession & $>640$ & Rumors regarding suicide bomber\\
2005 & Chennai, India & Disaster Area & 42 & Rush for flood relief supplies\\
2006 & Mena, Saudi Arabia & Jamarat Bridge & 363 & Overcrowding \\
2006  & Pilippines & Stadium & 79 & Rush for game show tickets \\
2006 & Ibb, Yemen & Stadium & 51 & Rally for Yemeni president\\
\hline
\end{tabular}
}\end{center}
\end{table}                                                                     

\subsection{Situations of ``Panic''}\label{situ}

Panic stampede is one of the most tragic collective
behaviors \cite{Hills,Canter,Exp,Coll1,Coll3,SocPsy,SocPsy1,SocPsy2,Coll2,SocPsy4},
as it often leads to the death of people who are either crushed or trampled down by others.
While this behavior may be comprehensible in life-threatening
situations like fires in crowded buildings \cite{Fire2,Fire1},
it is hard to understand in cases of a
rush for good seats at a pop concert \cite{Stadion}
or without any obvious reasons. Unfortunately, the frequency of such 
disasters is increasing (see Table \ref{Tabl}), as 
growing population densities combined with easier
transportation lead to greater mass events like
pop concerts, sport events, and demonstrations. Nevertheless,
systematic empirical studies of panic \cite{Exp,Exp1} are
rare \cite{Fire1,Coll1,Stadion},
and there is a scarcity of quantitative theories capable of
predicting crowd dynamics at extreme densities
\cite{KlMeyeWaSc00,EvacSim,Ebihara,FireSim4,Stil00,Ham}.
The following features appear to be typical 
\cite{HelFaVi00a,panic}:
\begin{enumerate}
\item In situations of escape panic, 
individuals are getting nervous,
i.e. they tend to develop blind actionism. 
\item People try to move considerably faster than normal
\cite{eng1}.
\item Individuals start pushing, and interactions among people 
become physical in nature.
\item Moving and, in particular, passing of a bottleneck  frequently
becomes incoordinated \cite{Exp}. 
\item At exits, jams are building up \cite{Exp}. Sometimes, intermittent flows or
arching and clogging are observed \cite{eng1}.
\item The physical interactions in jammed crowds 
add up and can cause dangerous pressures up to
4,500 Newtons per meter \cite{Fire2,calibrate},
which can bend steel barriers or tear down brick walls.
\item The strength and direction of the forces acting in large crowds
can suddenly change \cite{turb}, pushing people around in an uncontrollable
way. This may cause people to fall.
\item Escape is slowed down by fallen or injured people turning
into ``obstacles''.
\item People tend to show herding behavior, i.e., 
to do what other people do \cite{Fire1,SocPsy1}. 
\item Alternative exits are often overlooked or not efficiently used
in escape situations \cite{Fire2,Fire1}.
\end{enumerate}

\subsection{Force Model for Panicking Pedestrians}\label{Gran1}

Additional, physical interaction forces $\vec{f}_{\alpha\beta}^{\rm ph}$
come into play when pedestrians get so close to each other that they have physical contact
(i.e. $d_{\alpha\beta} < r_{\alpha\beta} = r_\alpha+r_\beta$, where $r_\alpha$ means the 
``radius'' of pedestrian $\alpha$) \cite{panic}. In this case, which is mainly relevant to panic situations,
we assume also a {\em ``body force''} $k(r_{\alpha\beta}-d_{\alpha\beta})
\, \vec{n}_{\alpha\beta}$ counteracting body compression
and a {\em ``sliding friction force''} $\kappa (r_{\alpha\beta} - d_{\alpha\beta})
\, \Delta v_{\beta\alpha}^t \, \vec{t}_{\alpha\beta}$ impeding {\em relative} tangential motion. 
Inspired by the formulas for granular interactions \cite{Herrmann,Granular}, we assume
\begin{equation}
\vec{f}_{\alpha\beta}^{\rm ph}(t) = k \Theta(r_{\alpha\beta}-d_{\alpha\beta}) 
\vec{n}_{\alpha\beta} + \kappa \Theta(r_{\alpha\beta}-d_{\alpha\beta}) 
\Delta v_{\beta\alpha}^t \, \vec{t}_{\alpha\beta} \, ,
\label{FORMEL}
\end{equation}
where the function $\Theta(z)$ is equal to its argument $z$, if $z\ge 0$, otherwise 0. 
Moreover, $\vec{t}_{\alpha\beta} = (-n_{\alpha\beta}^2, n_{\alpha\beta}^1)$ means the tangential
direction and $\Delta v_{\beta\alpha}^t = (\vec{v}_\beta -\vec{v}_\alpha) 
\cdot \vec{t}_{\alpha\beta}$
the tangential velocity difference, while $k$ and $\kappa$ represent large
constants.  (Strictly speaking, friction effects already set in before pedestrians
touch each other, because of the psychological tendency not to pass other individuals
with a high relative velocity, when the distance is small.) 
\par
The interactions with the boundaries of walls and other obstacles 
are treated analogously to pedestrian interactions, i.e., if
$d_{\alpha i}(t)$ means the distance to obstacle or boundary $i$,
$\vec{n}_{\alpha i}(t)$ denotes the direction perpendicular to it, and
$\vec{t}_{\alpha i}(t)$ the direction tangential to it,
the corresponding interaction force with the boundary reads
\begin{equation}
 \vec{f}_{\alpha i} = \left\{ A_\alpha \exp[(r_{\alpha}-d_{\alpha i})/B_\alpha]
 + k \Theta(r_\alpha-d_{\alpha i}) \right\} \vec{n}_{\alpha i} 
 - \kappa \Theta(r_\alpha-d_{\alpha i}) (\vec{v}_\alpha\cdot\vec{t}_{\alpha i}) 
 \, \vec{t}_{\alpha i} \, .
\end{equation}
\par
Finally, fire fronts are reflected by repulsive social forces similar those describing walls,
but they are much stronger. The physical interactions, however, are qualitatively different,
as people reached by the fire front become injured and immobile ($\vec{v}_\alpha = \vec{0}$).

\subsection{Collective Phenomena in Situations of ``Panic''} \label{Gran2}

Inspired by the observations discussed in Sec. \ref{situ},
we have simulated situations of ``panic'' escape in the computer, assuming the following
features:
\begin{enumerate}
\item People are getting nervous, resulting in a higher level of fluctuations.
\item They are trying to escape from the source of panic, which can be reflected 
by a significantly higher desired velocity $v_\alpha^0$.
\item Individuals in complex situations, who do not know what is the right thing to do, 
orient at the actions of their neighbours, i.e. they tend to do what other people do.
We will describe this by an additional herding interaction.
\end{enumerate}
We will now discuss the fundamental collective 
effects which fluctuations, increased desired velocities, and herding behavior can have
according to simulations. Note that, in contrast to other approaches, 
we do not assume or imply that individuals in
panic or emergency situations would behave relentless and asocial, although they sometimes do.

\subsubsection{Herding and ignorance of available exits:} 

\begin{figure}[htbp] 
\begin{center}
    \includegraphics[width=0.42\textwidth]{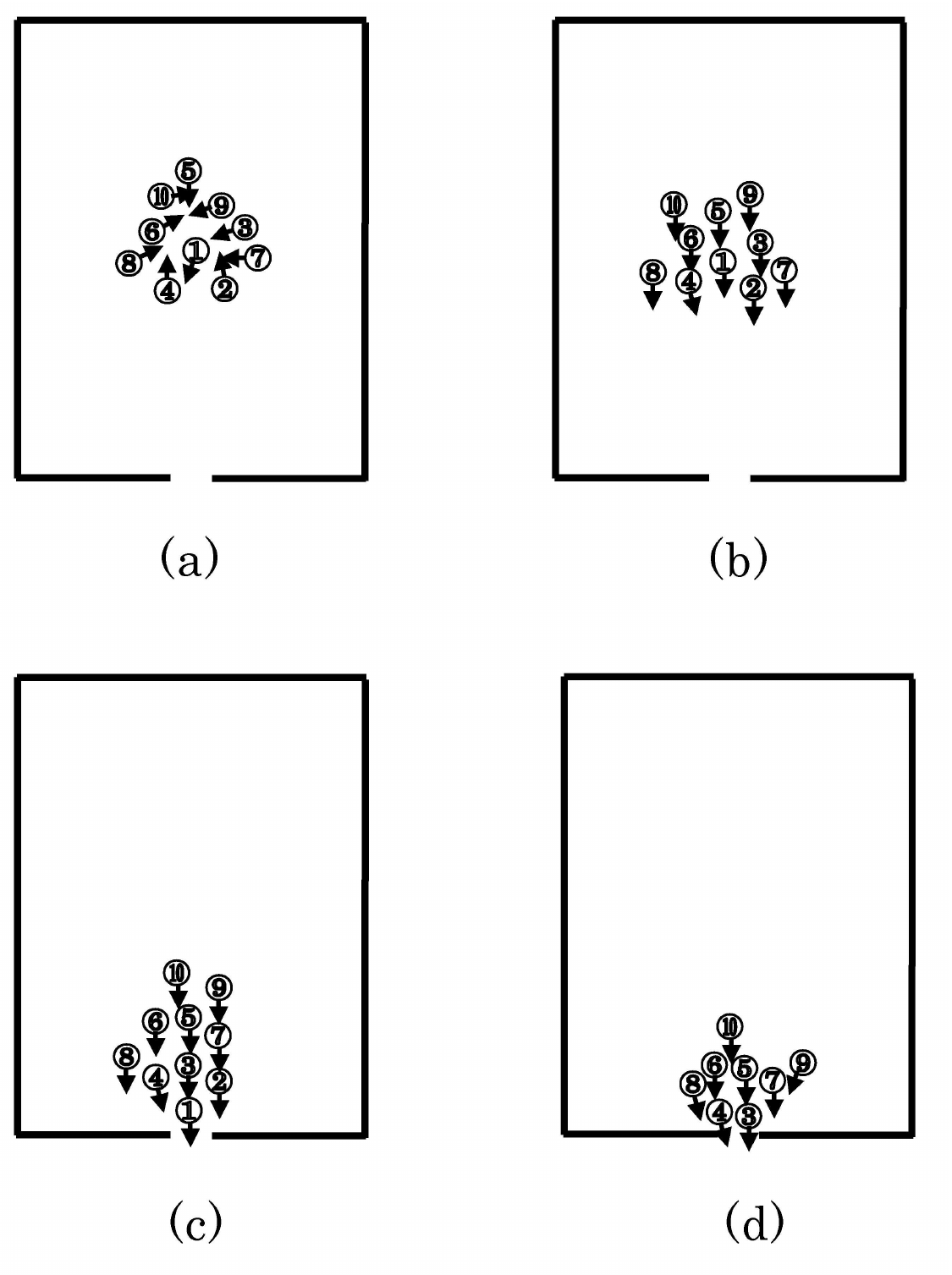}\hspace*{10mm}
    \includegraphics[width=0.42\textwidth]{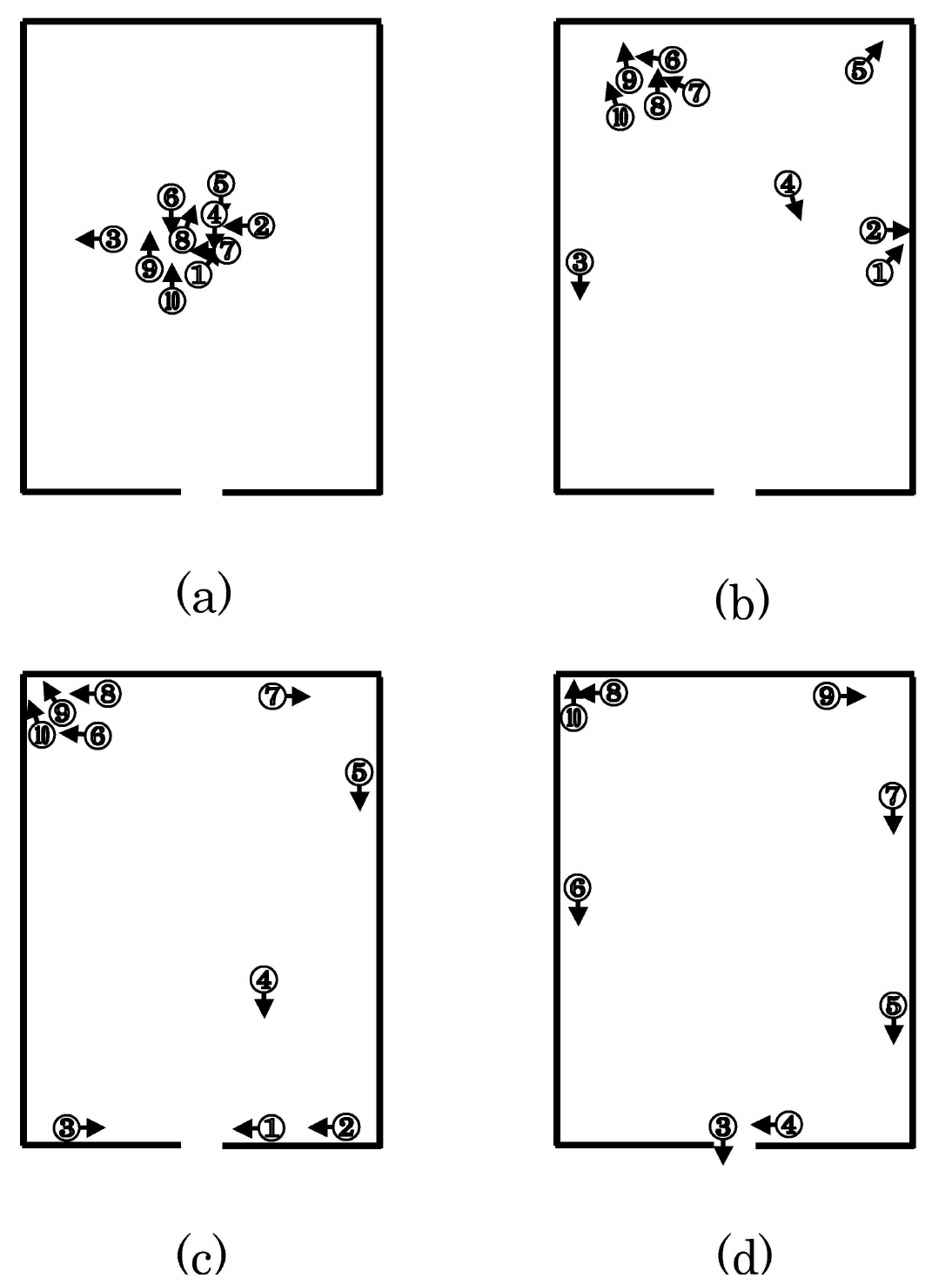}
\end{center}
\caption[]{Left: Normal leaving of a room, when the exit is well visible. Snapshots of a
video-recorded experiment with 10 people after (a) $t=0$ seconds (initial condition), (b) $t=1$ sec.,
(c) $t=3$ sec., and (d) $t=5$ seconds. The face directions are indicated by arrows.
Right: Escape from a room with no visibility, e.g. due to dense smoke or a power
blackout. Snapshots of an experiment with test persons, whose eyes were covered
by masks, after $t=0$ seconds (initial condition), $t=5$ sec., (c) $t=10$ sec., and (d)
$t=15$ seconds. (After Ref. \cite{Isobe}.)}
\label{Smoky}
\end{figure}

If people are not sure what is the best thing to do, there is a tendency to show
a ``herding behavior'', i.e. to imitate the behavior of others. 
Fashion, hypes and trends are examples for this. The phenomenon
is also known from stock markets, and particularly pronounced when people are
anxious. Such a situation is, for example, given if people need to escape from a
smoky room. There, the evacuation dynamics is very different from normal leaving
(see Fig.~\ref{Smoky}).
\par
Under normal visibility, everybody easily finds an exit and uses more or less the shortest
path. However, when the exit cannot be seen, evacuation is much less efficient and may take a long time.
Most people tend to walk relatively straight into the direction in which they suspect an exit, but
in most cases, they end up at a wall. Then, they usually move along it in one of the two possible directions, 
until they finally find an exit \cite{Isobe}. If they
encounter others, there is a tendency to take a decision for one direction and move collectively.
Also in case of accoustic signals, people may be attracted into the same direction.
This can lead to over-crowded exits, while other exits are ignored. The same can happen
even for normal visibility, when people are not well familiar with their environment 
and are not aware of the directions of the emergency exits.
\par
Computer simulations suggest that neither
individualistic nor herding behavior performs well \cite{panic}.
Pure individualistic behavior means that each
pedestrian finds an exit only accidentally, while pure herding
behavior implies that the complete crowd is eventually moving into
the same and probably congested
direction, so that available emergency exits are not efficiently used. Optimal 
chances of survival are expected for a certain
mixture of individualistic and herding behavior, where 
individualism allows {\em some} people to detect the exits and herding guarantees
that successful solutions are imitated by small groups of others \cite{panic}.

\subsubsection{``Freezing by heating'':}

\begin{figure}[htbp]
\unitlength1cm
\begin{center}
\includegraphics[width=8\unitlength]{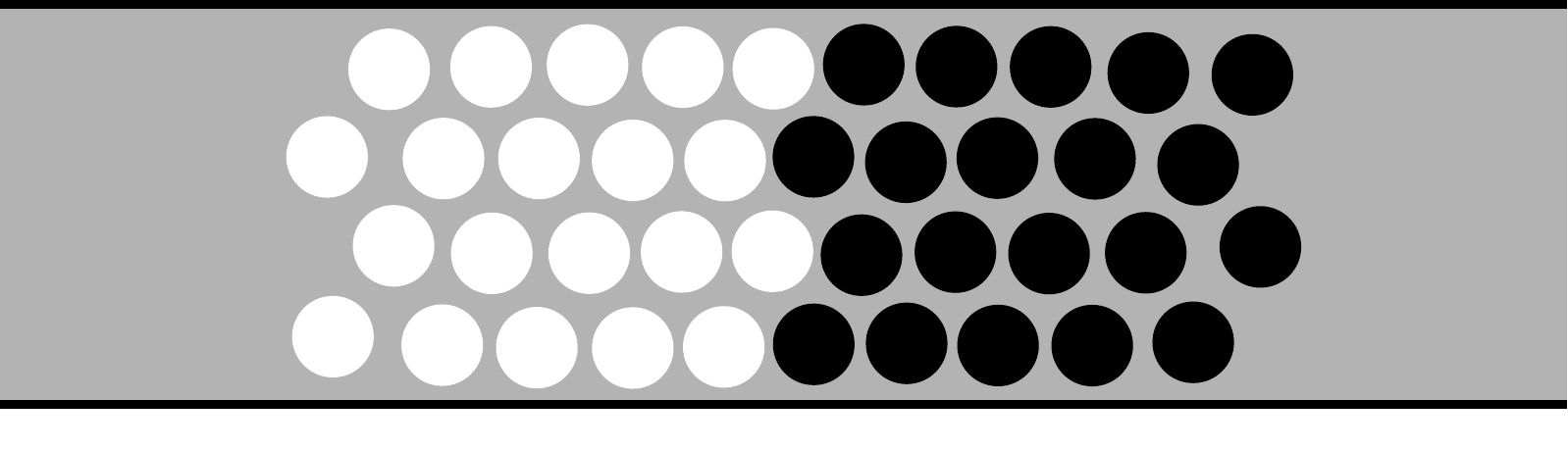}
\vspace*{-5mm}
\end{center}
\caption[]{Result of the noise-induced formation of a ``frozen'' state in a (periodic)
corridor used by oppositely moving pedestrians 
(after Ref. \cite{HelFaVi00a}).\label{FREEZ}\label{freeze}}
\end{figure}

Another effect of getting nervous has been investigated in Ref.~\cite{HelFaVi00a}.
Let us assume the individual fluctuation strength, i.e. the standard deviation of
the noise term $\vec{\xi}_\alpha$, is given by
\begin{equation}
  \eta_\alpha = (1 - n_\alpha) \eta_0 + n_\alpha \eta_{\rm max} \, ,
\end{equation}
where $n_\alpha$ with $0 \le n_\alpha \le 1$ measures the nervousness of pedestrian
$\alpha$. The parameter $\eta_0$ means 
the normal and $\eta_{\rm max}$ the maximum fluctuation strength.
It turns out that, at sufficiently high pedestrian densities, 
lanes are destroyed by increasing the fluctuation strength (which is
analogous to the temperature). However, instead of the expected transition 
from the ``fluid''  lane state to a disordered, ``gaseous''
state, a ``solid'' state is formed. 
It is characterized by a blocked, ``frozen'' situation so that one
calls this paradoxial transition {\em ``freezing
by heating''} (see Fig.~\ref{freeze}). Notably enough, the blocked state has
a {\em higher} degree of order, although the internal
energy is {\em increased} \cite{HelFaVi00a}. 
\par
The preconditions for this unusual freezing-by-heating
transition are the driving term $v_\alpha^0 \vec{e}_\alpha^0/\tau_\alpha$ and the dissipative friction
$- \vec{v}_\alpha/\tau_\alpha$, while the sliding friction force is not required. Inhomogeneities in the 
channel diameter or other impurities which temporarily slow down pedestrians
can further this transition at the respective places. Finally note that a transition from
fluid to blocked pedestrian counter flows is also observed, when a critical density
is exceeded, as impatient pedestrians enter temporary gaps in the opposite lane to overtake others \cite{MurIrNa99,HelFaVi00a}. However, in contrast to computer simulations, resulting deadlocks are usually
not permanent in real crowds, as turning the bodies (shoulders) often allows pedestrians 
to get out of the blocked area.

\subsubsection{Intermittent flows, faster-is-slower effect, and ``phantom panic'':}

If the overall flow towards a bottleneck is higher than the overall outflow from it, 
a pedestrian queue emerges \cite{PRL}. In other words, a
waiting crowd is formed upstream of the bottleneck. High densities can result,
if people keep heading forward, as this eventually leads to higher and higher compressions.
Particularly critical situations may occur if the arrival flow is much higher than the departure flow,
especially if people are trying to get towards a strongly desired goal (``aquisitive panic'') or
away from a perceived source of danger (``escape panic'') with an increased driving force
$v_\alpha^0\vec{e}_\alpha^0/\tau$. In such situations, the high
density causes coordination problems, as several people compete for the same few gaps.
This typically causes body interactions and frictional effects, which can slow down 
crowd motion or evacuation {\em (``faster is slower effect'')}. 
\par
\begin{figure}[htbp] 
\unitlength1cm
\begin{center}
\begin{picture}(11,15)
\put(0,8.6){\includegraphics[width=0.84\textwidth]{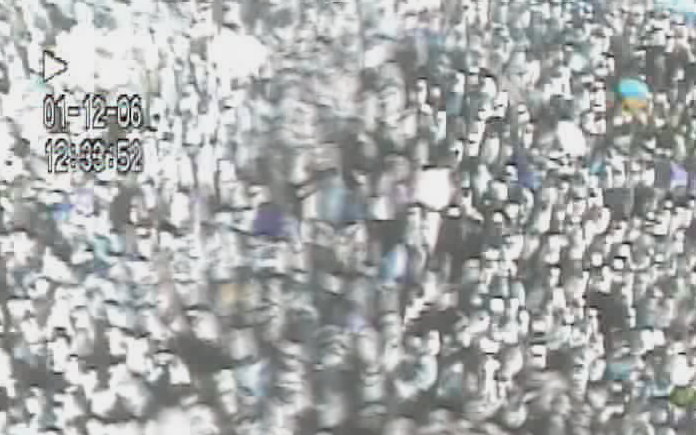}}
\put(-0.8,4.7){\includegraphics[width=1\textwidth]{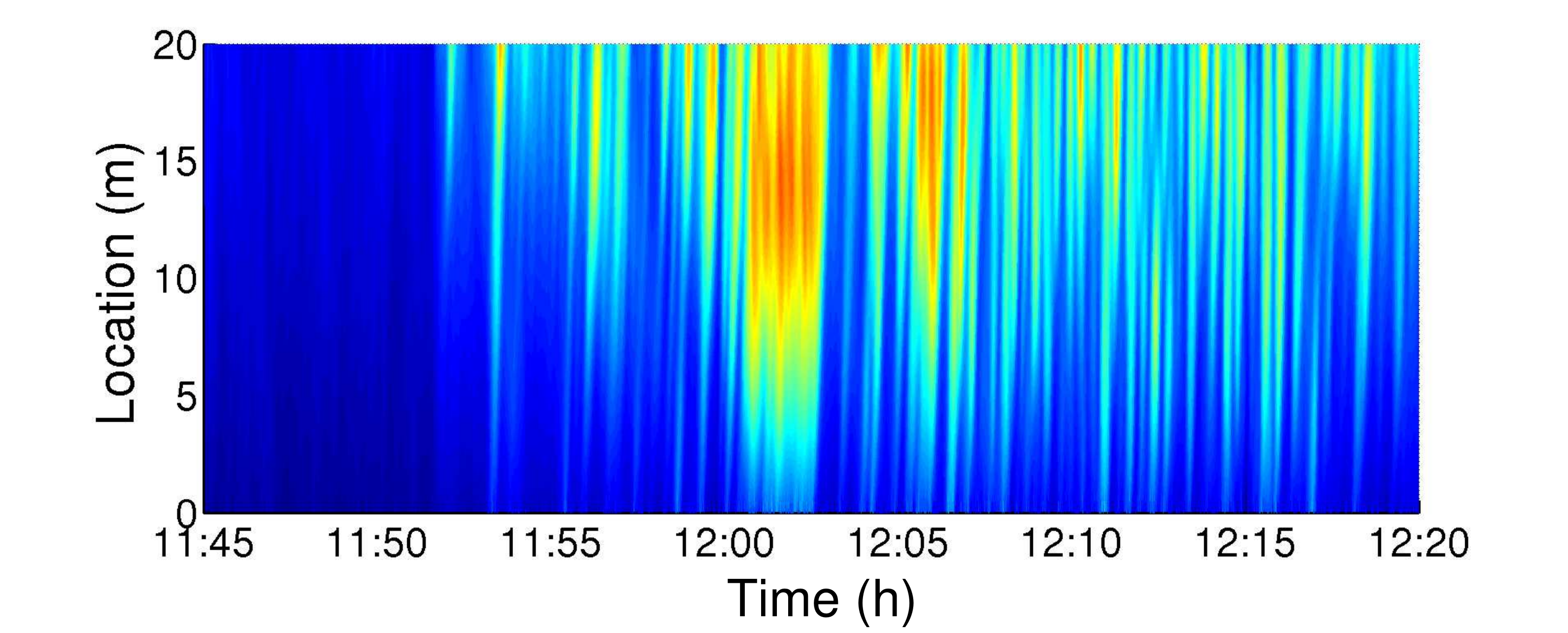}}
\put(3.5,-0.7){\includegraphics[width=0.6\textwidth]{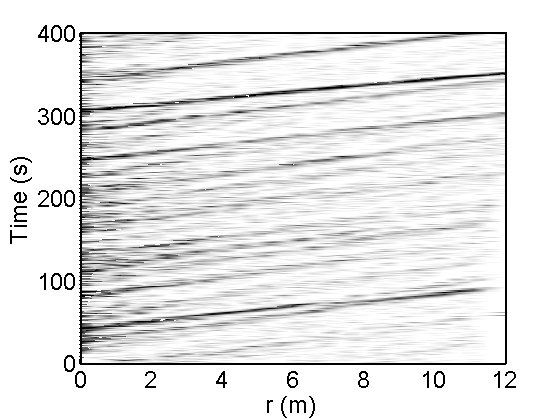}}
\put(-0.3,-0.2){\includegraphics[width=0.4\textwidth,angle=90]{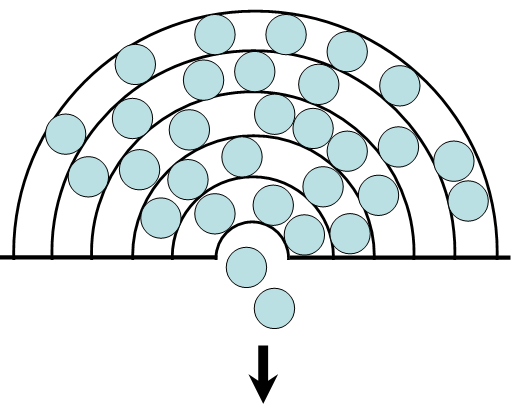}}
\end{picture}
\end{center}
\caption[]{Top: Long-term photograph showing stop-and-go waves in a densely packed street.
While stopped people appear relatively sharp, people moving from right to left 
have a fuzzy appearance. Note that gaps propagate from left to right. Middle:
Empirically observed stop-and-go waves in front of the entrance to the Jamarat Bridge on 
January 12, 2006 (after \cite{turb}), where pilgrims moved from left to right. Dark areas
correspond to phases of motion, light colors to stop phases. The ``location''  coordinate
represents the  distance to the beginning of the narrowing, i.e. to the cross section of reduced width. Bottom left:
Illustration of the ``shell model'' (see Ref. \cite{PRL}), in particular of situations where several pedestrians compete
for the same gap, which causes coordination problems. Bottom right: Simulation results of
the shell model. The observed stop-and-go waves 
result from the alternation of forward pedestrian motion and backward gap propagation.}
\label{stop}
\end{figure}

A possible consequence of these coordination problems are intermittent flows. In such
cases, the outflow from the bottleneck is not constant, but it is typically interrupted.
While one possible origin of the intermittent flows are clogging and arching effects
as known from granular flows through funnels or hoppers \cite{Herrmann,Granular},
stop-and-go waves have also been observed in more than 10 meter wide streets and in
the 44 meters wide entrance area to the Jamarat Bridge during the pilgrimage in January 12, 2006
\cite{turb}, see Fig. \ref{stop}. Therefore, it seems to be important that people do not move continuously, but have
minimum strides \cite{Hel91}. That is, once a person is stopped, he or she will not
move until some space opens up in front. However, increasing impatience will eventually
reduce the minimum stride, so that people eventually start moving again, even if the outflow
through the bottleneck is stopped. This will lead to a further compression of the crowd.
\par 
In the worst case, such behavior can trigger a {\em ``phantom panic''}, i.e.
a crowd disaster {\em without} any serious reasons (e.g., in Moscow, 1982).  
For example, due to the ``faster-is-slower effect''
panic can be triggered by small pedestrian counterflows \cite{Fire2}, 
which cause delays to the crowd
intending to leave. Consequently, stopped pedestrians in the back, who do not see the reason
for the temporary slowdown, are
getting impatient and pushy. In accordance with observations
\cite{Hel91,Hel97a}, one may model this by increasing the desired velocity, for example,
by the formula
\begin{equation}
 v_\alpha^0(t) = [1-n_\alpha(t)]v_\alpha^0(0) + n_\alpha(t) v_\alpha^{\rm max}  \, .
\end{equation}
Herein, $v_\alpha^{\rm max}$ is
the maximum desired velocity and $v_\alpha^0(0)$ the initial one, corresponding to the
expected velocity of leaving. The time-dependent parameter
\begin{equation}
 n_\alpha(t) = 1 - \frac{\overline{v}_\alpha(t)}{v_\alpha^{\rm 0}(0)} 
\end{equation}
reflects the nervousness,  where $\overline{v}_\alpha(t)$
denotes the average speed into the desired direction of motion. 
Altogether, long waiting times increase the desired speed $v_\alpha^0$ or
driving force $v_\alpha^0(t)\vec{e}_\alpha^0/\tau$,
which can produce high densities and inefficient motion. This further increases
the waiting times, and so on, so that this tragic feedback can
eventually trigger so high pressures that people are crushed 
or falling and trampled. It is, therefore, imperative, to have sufficiently
wide exits and to prevent counterflows, when big crowds want to 
leave \cite{panic}.

\subsubsection{Transition to stop-and-go waves:} \label{sec_stopandgo}

Recent empirical studies of pilgrim flows in the area of Makkah, Saudi Arabia,
have shown that intermittent flows occur not only when bottlenecks are obvious.
On January 12, 2006, pronounced stop-and-go waves have been even observed upstream of 
the 44 meter wide entrance to the Jamarat Bridge \cite{turb}. While the pilgrim flows were smooth 
and continuous (``laminar'') over many hours, at 11:53am stop-and-go waves suddenly appeared and
propagated over distances of more than 30 meters (see Fig. \ref{stop}). The sudden transition
was related to a significant drop of the flow, i.e. with the onset of congestion \cite{turb}.
Once the stop-and-go waves set in, they persisted over more than 20 minutes.
\par
This phenomenon can be reproduced by a recent model based on two continuity equations, one
for forward pedestrian motion and another one for backward gap propagation \cite{PRL}. 
The model was derived from a ``shell model'' (see Fig. \ref{stop}) and describes 
very well the observed alternation between backward gap propagation and forward pedestrian motion.

\subsubsection{Transition to ``crowd turbulence'':}

\begin{figure}[htbp]
\unitlength1.2cm
\begin{center}
\begin{picture}(12,3.5)
\put(0.5,0){\includegraphics[height=4cm]{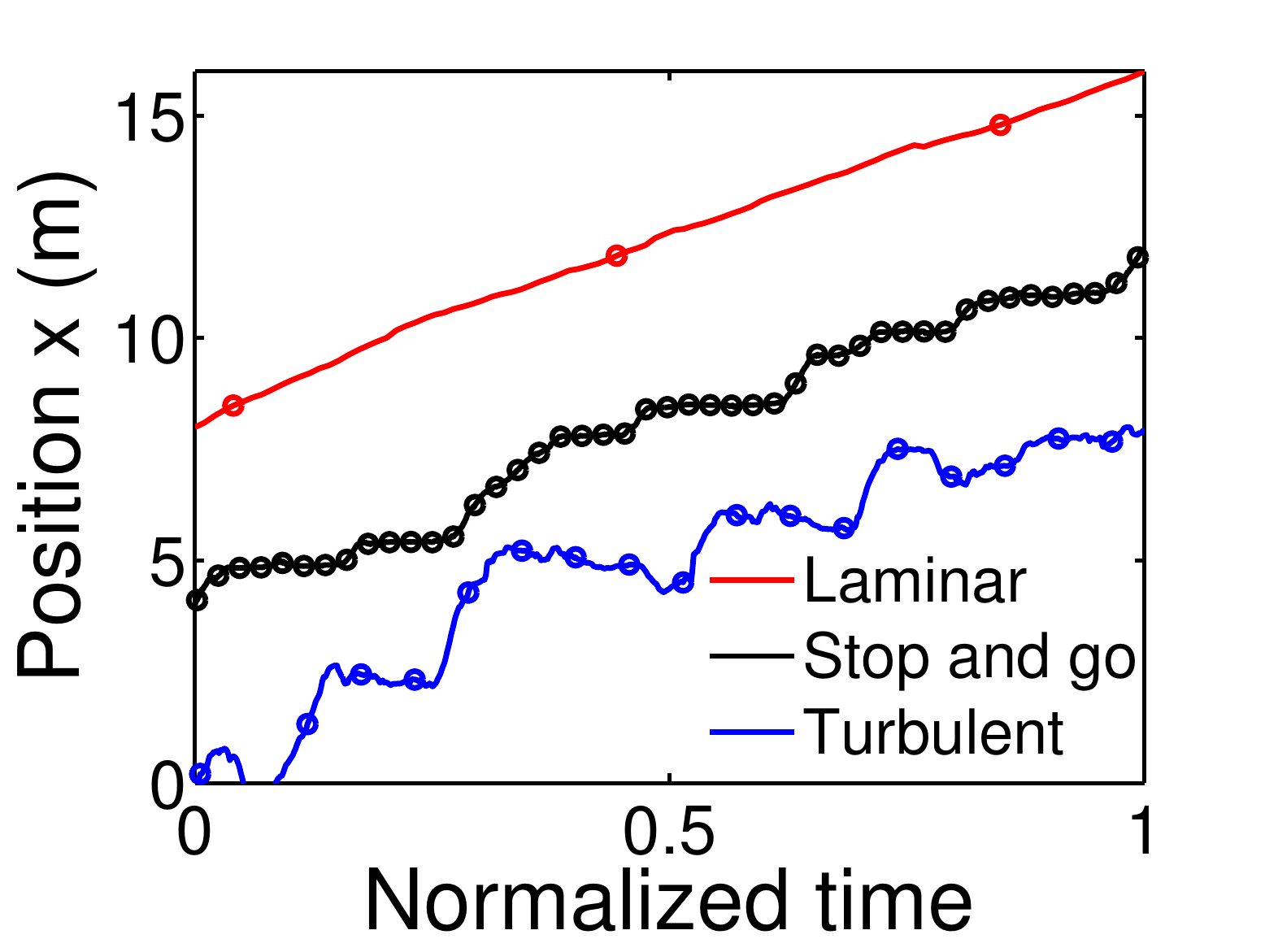}}
\put(5.3,-0.2){\includegraphics[height=4.2cm]{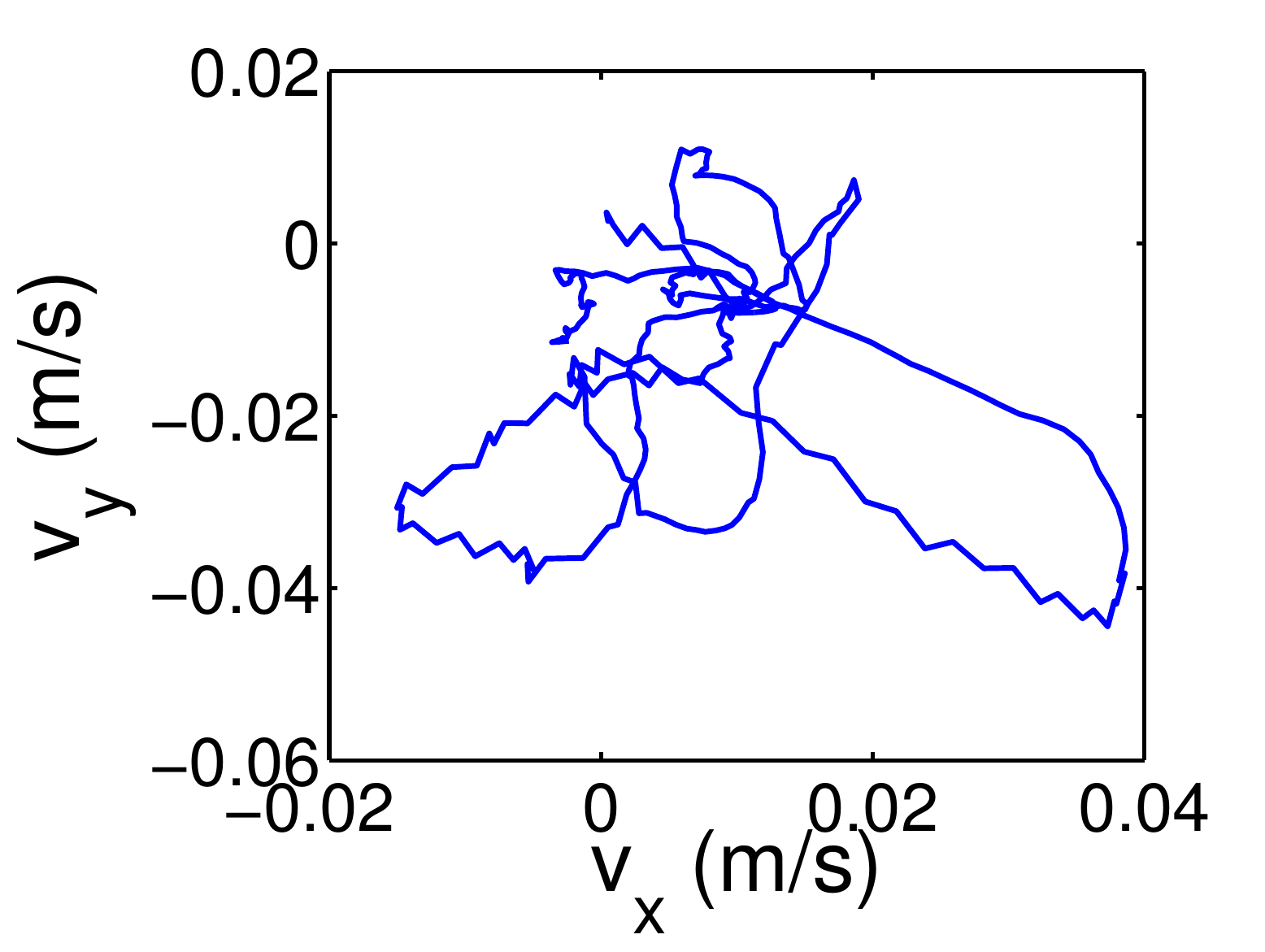}}
\end{picture}
\end{center} 
\caption{Pedestrian dynamics at different densities. 
Left: Representative trajectories (space-time plots) of pedestrians during the laminar, stop-and-go, and
turbulent flow regime. Each trajectory extends over a range of 8 meters, while
the time required for this stretch is normalized to 1. To indicate the different speeds, symbols are included
in the curves every 5 seconds. While the laminar flow (top line) is fast and smooth, motion is temporarily interrupted
in stop-and-go flow (medium line), and backward motion can occur in ``turbulent'' flows (bottom line).
Right: Example of the temporal evolution of the velocity components $v_x(t)$ into the average direction of motion
and $v_y(t)$ perpendicular to it in ``turbulent flow'', which occurs when the crowd density is extreme.
One can clearly see the irregular motion into all possible directions characterizing
``crowd turbulence''. For details see Ref.~\cite{turb}.}
\label{turbu}
\end{figure}

On the same day, around 12:19, the density reached even higher values and 
the video recordings showed a sudden transition from stop-and-go waves
to {\em irregular} flows  (see Fig.~\ref{turbu}). These irregular flows were characterized
by random, unintended displacements into all possible directions, which pushed people 
around. With a certain likelihood, this caused them to stumble. 
As the people behind were moved by the crowd as well and could not stop, fallen 
individuals were trampled, if they did not get back on their feet quickly enough. 
Tragically, the area of trampled people grew more and more in the course of time, 
as the fallen pilgrims became obstacles for others \cite{turb}. The result was one of the
biggest crowd disasters in the history of pilgrimage.
\par
How can we understand this transition to irregular crowd motion?
A closer look at video recordings of the crowd reveals that, 
at this time, people were so densely packed that they  
were moved involuntarily by the crowd. This
is reflected by random displacements into all possible directions. 
To distinguish these irregular flows
from laminar and stop-and-go flows and due to their visual appearance, 
we will refer to them as {\em ``crowd turbulence''}. 
\par
As in certain kinds of fluid flows, 
``turbulence'' in crowds results from a sequence of instabilities 
in the flow pattern. Additionally, one finds a sharply peaked probability density function
of velocity increments 
\begin{equation}
 V_x^\tau = V_x(\vec{r},t+\tau) -V_x(\vec{r},t) \, ,
\end{equation}
which is typical for turbulence \cite{exchange}, if the time shift $\tau$ is small enough \cite{turb}.
One also observes a power-law scaling of the displacements indicating self-similar behaviour \cite{turb}. 
As large eddies are not detected, however, the similarity 
with {\em fluid} turbulence is limited, but there is still an analogy to 
turbulence at currency exchange markets \cite{exchange}. 
Instead of vortex cascades like in turbulent fluids, one rather finds a hierarchical fragmentation dynamics:
At extreme densities, individual motion is replaced by mass motion, but there is a stick-slip
instability which leads to ``rupture'' when the stress in the crowd becomes too large. That is,
the mass splits up into clusters of different sizes with strong velocity correlations {\em inside} and
distance-dependent correlations {\em between} the clusters. 
\par
``Crowd turbulence'' has further specific features \cite{turb}.
Due to the physical contacts among people in extremely dense crowds, we expect
commonalities with granular media.  In fact, dense driven granular media 
may form density waves, while moving forward \cite{Peng}, and can display 
turbulent-like states \cite{turb1,turb2}. Moreover, under quasi-static conditions \cite{turb1}, 
force chains \cite{fragile} are building up, causing strong 
variations in the strengths and directions of local forces. As in 
earthquakes \cite{earthquake,earthquake2} this can lead to events 
of sudden, uncontrollable stress release with power-law distributed displacements. 
Such a power-law has also been discovered by video-based crowd analysis \cite{turb}.

\subsection{Some Warning Signs of Critical Crowd Conditions}

\begin{figure}[htbp] 
\begin{center}
    \hspace*{2mm}\includegraphics[width=0.48\textwidth]{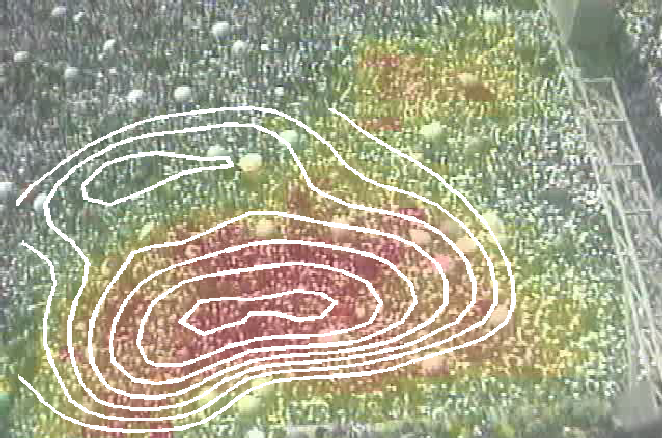}\hspace*{6mm}
    \includegraphics[width=0.45\textwidth]{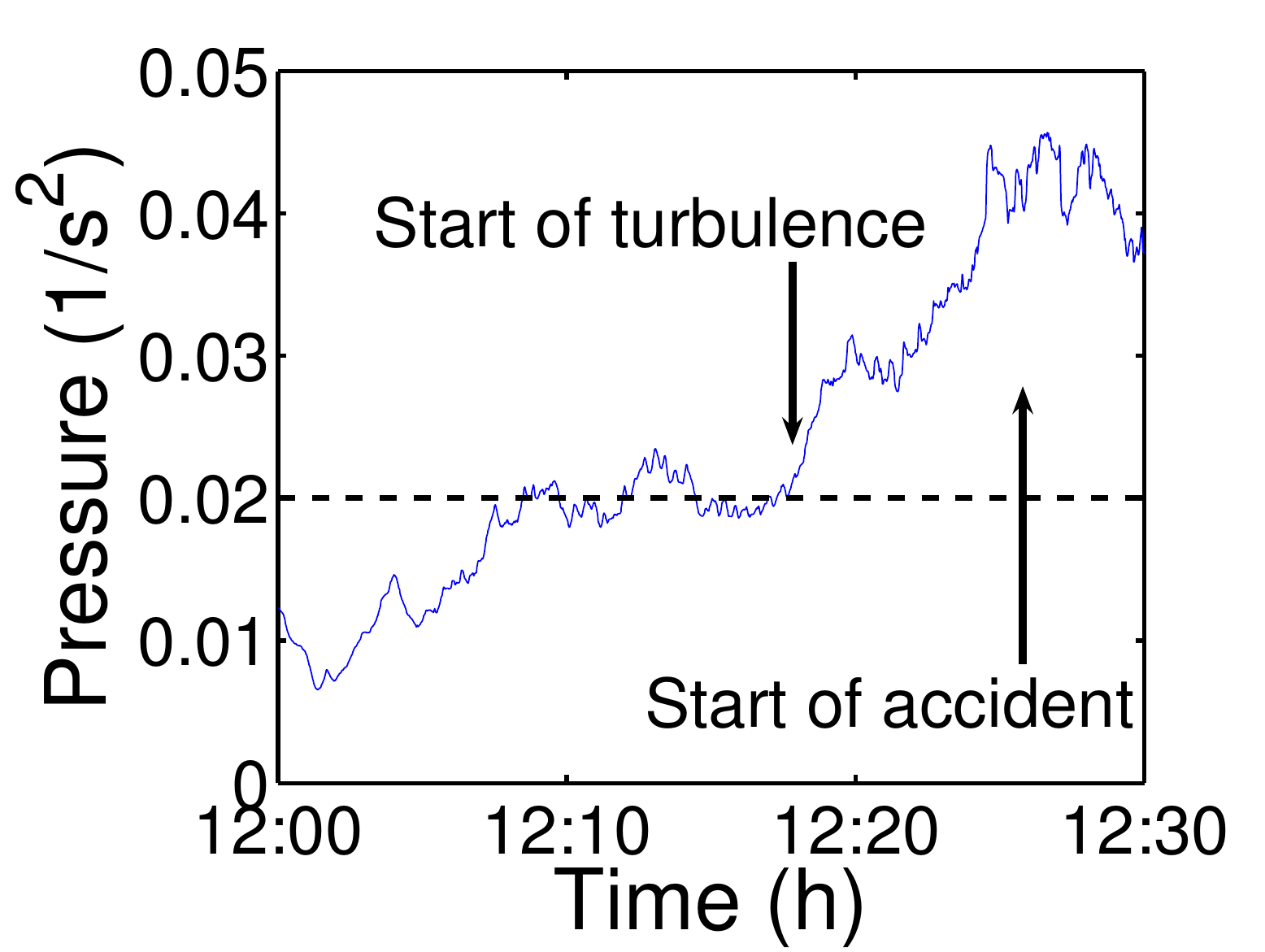}
\end{center}
\caption[]{Left: Snapshot of the on-line visualization of ``crowd pressure''. 
Red colors (see the lower ellipses) indicate areas of critical crowd conditions. 
In fact, the sad crowd disaster during the Muslim pilgrimage on January 12, 2006, started
in this area. Right: The ``crowd pressure'' is a quantitative measure of the onset of ``crowd turbulence''.
The crowd disaster started when the ``crowd pressure'' reached particularly high values. For details see Ref.~\cite{turb}.}
\label{turbul}
\end{figure}
Turbulent waves are experienced in dozens of crowd-intensive events each year
all over the world \cite{FRUIN}. Therefore, it is necessary to understand why, where and
when potentially critical situations occur.
Viewing real-time video recordings is not very suited to identify critical crowd conditions: 
While the average density rarely exceeds
values of 6 persons per square meter, the local densities can reach almost twice as large values
\cite{turb}. It has been found, however, that even evaluating the local densities 
is not enough to identify the critical times and locations precisely, 
which also applies to an analysis of the 
velocity field \cite{turb}. The decisive quantity is rather the ``crowd pressure'',
i.e. the density, multiplied with the variance of speeds.
It allows one to identify critical locations 
and times (see Fig. \ref{turbul}). 
\par
There are even advance warning signs of critical crowd conditions:
The crowd accident on January 12, 2006 started about 10 minutes 
after ``turbulent'' crowd motion set in, i.e. after the ``pressure'' exceeded a value of 0.02/s$^2$
(see Fig. \ref{turbul}). Moreover, it occured more than 30 minutes after 
stop-and-go waves set in, which can be easily detected in accelerated surveillance videos.
Such advance warning signs of critical crowd conditions 
can be evaluated on-line by an automated video analysis system. In many cases, this can 
help one to gain time for corrective measures like flow control, 
pressure-relief strategies, or the separation of crowds into blocks to stop the propagation
of shockwaves \cite{turb}. 
Such anticipative crowd control could increase the level of safety during 
future mass events.

\subsection{Evolutionary Optimization of Pedestrian Facilities}

Having understood some of the main factors causing crowd disasters, it is interesting
to ask how pedestrian facilities can be designed in a way that maximizes the efficiency of
pedestrian flows and the level of safety. One of the major goals during mass events must
be to avoid extreme densities. These often result from the onset of congestion at bottlenecks,
which is a consequence of the breakdown of free flow and causes an increasing degree
of compression. When a certain critical density is increased (which depends on the size distribution
of people), this potentially implies high
pressures in the crowd, particularly if people are impatient due to long delays or panic.
\par
The danger of an onset of congestion can
be minimized by avoiding bottlenecks.
Notice, however, that jamming can also
occur at widenings of escape routes \cite{panic}.
This surprising fact results from disturbances due to pedestrians, 
who try to overtake each other and expand in the wider area because of their repulsive
interactions.  These squeeze into the main stream again at the end of the 
widening, which acts like a bottleneck and leads to jamming.
The corresponding drop of efficiency $E$ is more pronounced,
\begin{enumerate}
\item if the corridor is narrow, 
\item if the pedestrians have different or high desired velocities, and 
\item if the pedestrian density in the corridor is high. 
\end{enumerate}
\par
Obviously, the emerging pedestrian flows decisively depend on 
the geometry of the boundaries. 
They can be simulated on a computer already in the planning phase of
pedestrian facilities. Their configuration
and shape can be systematically varied,
e.g. by means of evolutionary algorithms 
\cite{Bolay,Baeck} and evaluated on the
basis of particular mathematical performance measures 
\cite{Hel97a}. Apart from the
{\em efficiency} 
\begin{equation}
 E = \frac{1}{N} \sum_{\alpha} 
 \frac{\vec{v}_\alpha \cdot \vec{e}_\alpha^0}
  {v_\alpha^0} \label{EFF}
\end{equation}
we can, for example, define the {\em measure of comfort} $C = (1-D)$
via the discomfort 
\begin{equation}
 D = \frac{1}{N} \sum_{\alpha} \frac{\,\overline{(\vec{v}_\alpha
 - \overline{\vec{v}_\alpha})^2}\,}{\overline{(\vec{v}_\alpha)^2}}
 = \frac{1}{N} \sum_{\alpha} \left( 1 - \frac{\overline{\vec{v}_\alpha}^2}
 {\,\overline{(\vec{v}_\alpha)^2}\,} \right) \, .
\end{equation}
The latter is again between 0 and 1 and
reflects the frequency and degree of sudden velocity changes, i.e. 
the level of discontinuity of walking due to necessary avoidance maneuvers. 
Hence, the optimal
configuration regarding the pedestrian requirements is the one
with the highest values of efficiency and comfort. 
\par\begin{figure} \begin{center}
	\includegraphics[width=0.45\textwidth]{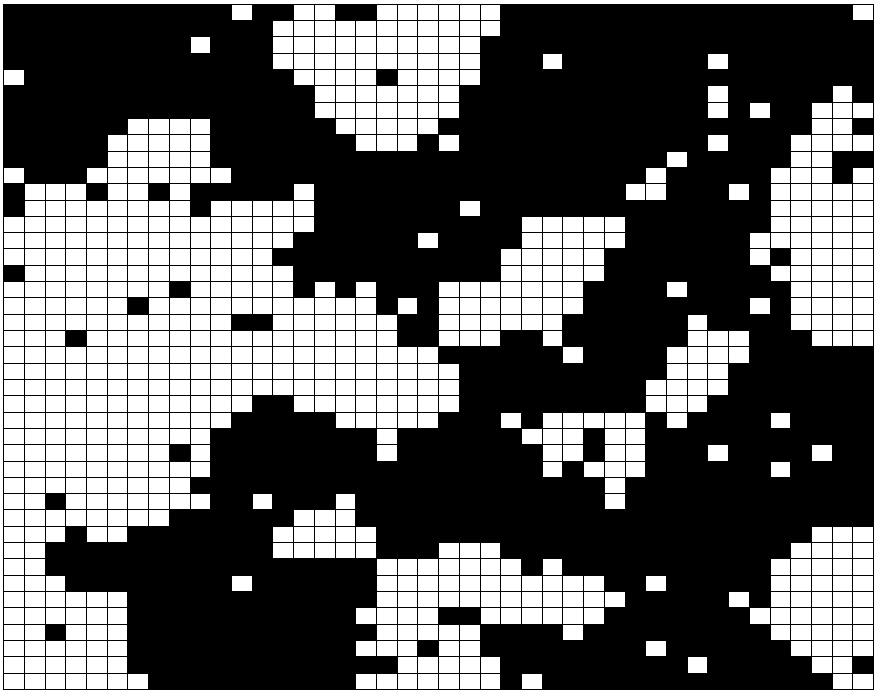}\hspace*{7mm}
	\includegraphics[width=0.45\textwidth]{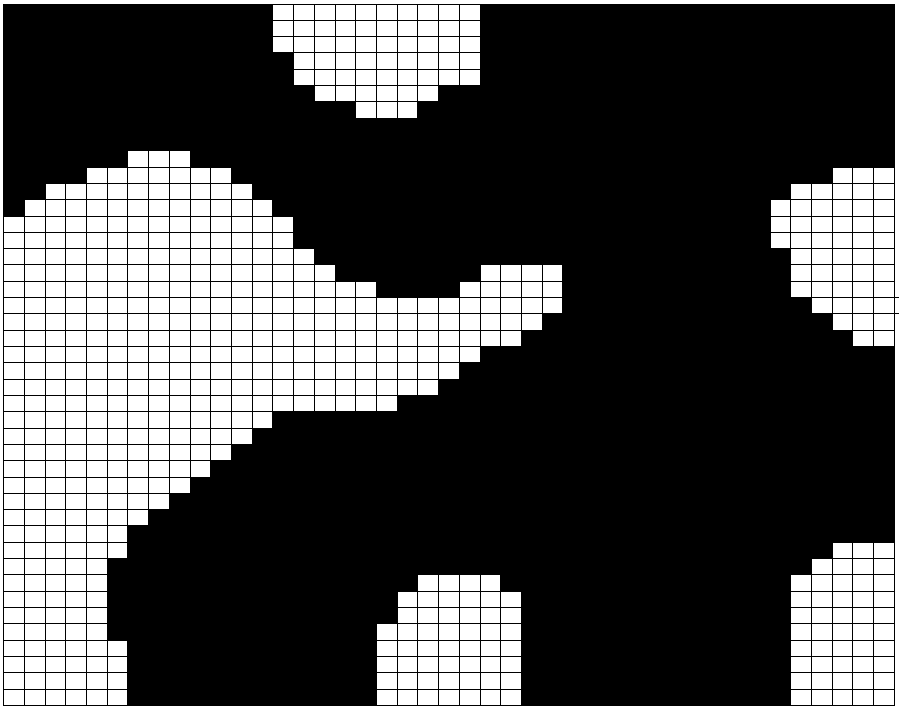}
\caption{The evolutionary optimization based on Boolean grids 
uses a two-stage algorithm (see Ref. \cite{Boolean} for details). Left: In the ``randomization stage'', obstacles are 
distributed over the grid with some randomness, thereby allowing for 
the generation and testing of new topologies (architectures). Right: In the ``agglomeration stage'', 
small nearby obstacles are clustered to form larger objects with smooth
boundaries. After several iterations, the best performing designs are reasonably shaped.
See Fig. \ref{designs} for examples of possible bottleneck designs.}
\label{Bool}
\end{center} \end{figure}
\begin{figure} \begin{center}
	\includegraphics[width=0.45\textwidth]{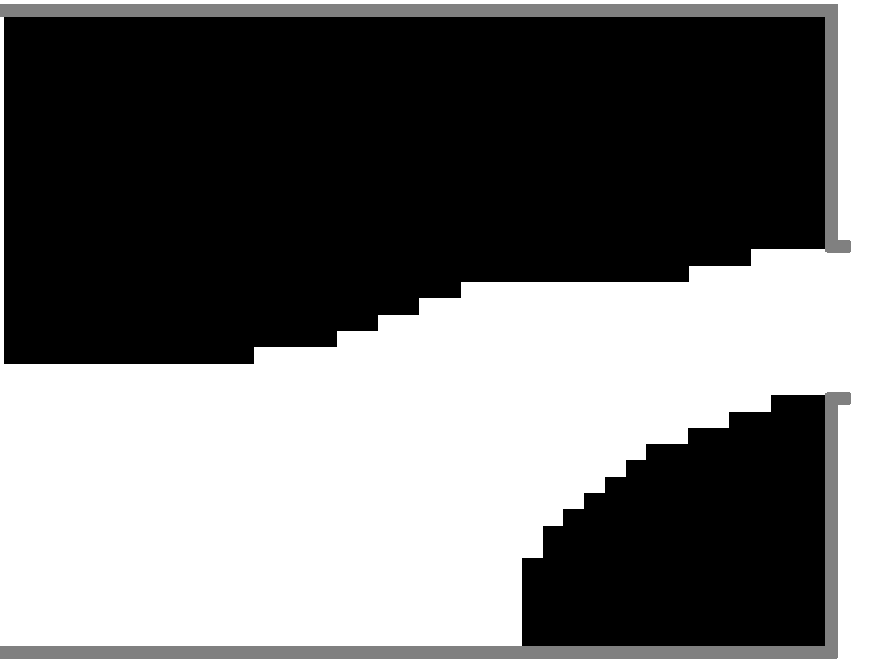}\hspace*{5mm}
	\includegraphics[width=0.45\textwidth]{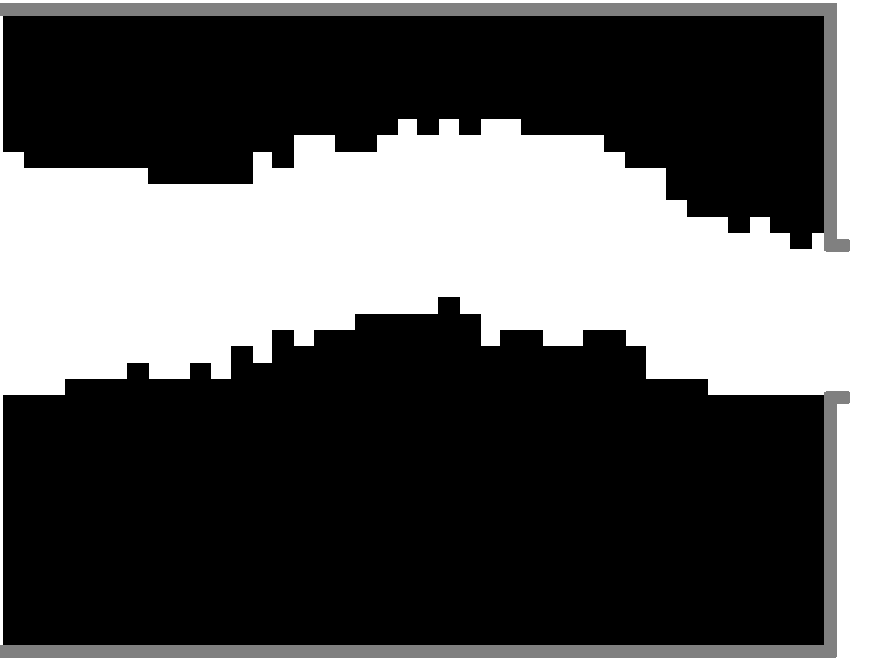}
\end{center} 
\caption{Two examples of improved designs for cases with a bottleneck along the escape
route of a large crowd, obtained with an evolutionary algorithm based on Boolean grids
(after Ref. \cite{Boolean}). People were assumed to move from left to right only. 
Left: Funnel-shaped escape route. Right:  Zig-zag design.}
\label{designs}
\end{figure}
\begin{figure}[htbp] 
\begin{center}
    \includegraphics[width=0.52\textwidth]{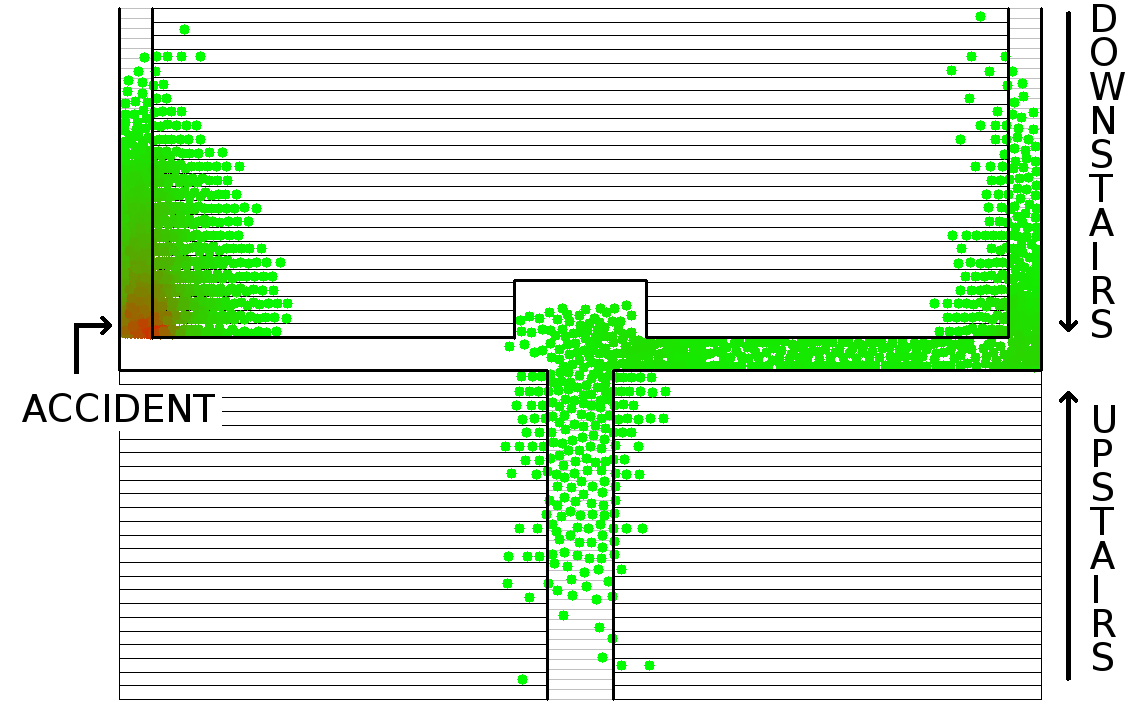}
    \includegraphics[width=0.47\textwidth]{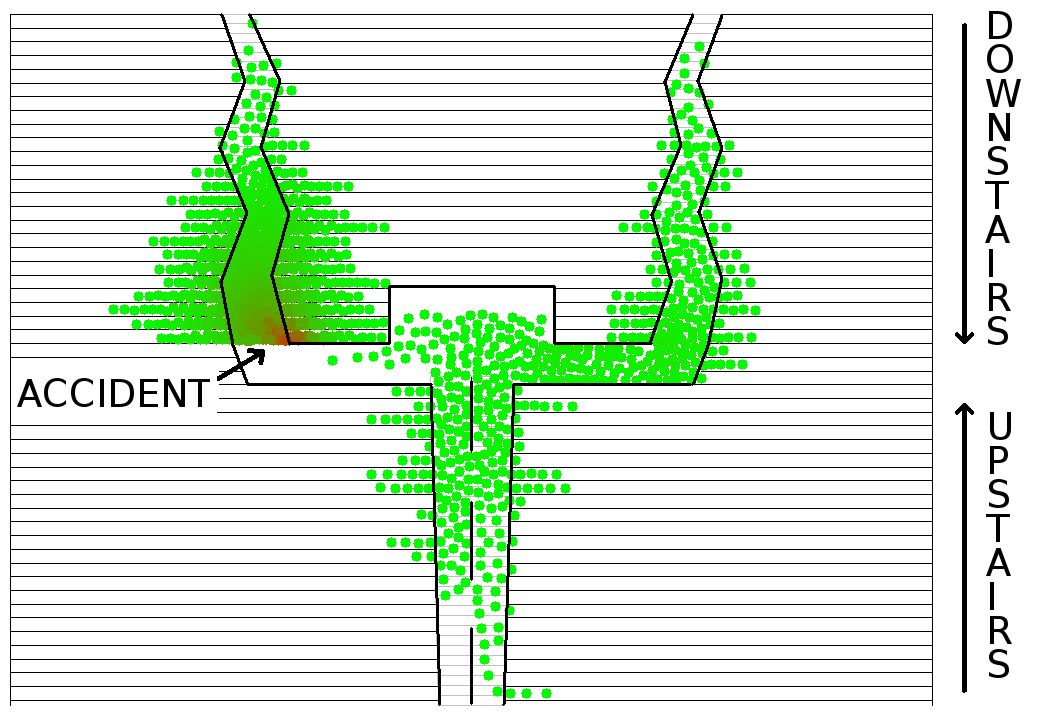}
\end{center}
\caption[]{Left: Conventional design of a stadium exit in an emergency 
scenario, where we assume that some pedestrians have fallen 
at the end of the downwards staircase to the left. The dark color indicates 
high pressures, since pedestrians are impatient and pushing from behind.
Right: In the improved design, the increasing diameter of corridors can 
reduce waiting times and impatience (even with the same number of seats),
thereby accelerating evacuation. Moreover, the zigzag design of the downwards 
staircases changes the pushing direction in the crowd. Computer simulations indicate
that the zig-zag design can reduce the average pressure in the crowd at the location of 
the incident by a factor of two. (After Ref. \cite{TranSci}.)}
\label{zigzag}
\end{figure}

During the optimization procedure, some or all of the following 
can be varied:
\begin{enumerate}
\item the location and form of planned buildings,
\item the arrangement of walkways, entrances, exits, staircases, 
elevators, escalators, and corridors,
\item the shape of rooms, corridors, entrances, and exits,
\item the function and time schedule. 
(Recreation rooms
or restaurants are often continuously frequented, rooms for conferences
or special events are mainly visited and left at peak periods, exhibition
rooms or rooms for festivities require additional 
space for people standing around, and some
areas are claimed by queues or through traffic.)
\end{enumerate}
In contrast to early evolutionary optimization methods, recent approaches allow
to change not only the dimensions of the different elements of pedestrian facilities,
but also to vary their topology. The procedure of such algorithms is illustrated in
Fig.~\ref{Bool}. Highly performing designs are illustrated in Fig.~\ref{designs}.
It turns out that, for an emergency evacuation route, it is favorable if the 
crowd does not move completely straight towards a bottleneck. 
For example, a zigzag design of the evacuation route
can reduce the pressure on the crowd upstream of a bottleneck (see Fig. \ref{zigzag}). 
The proposed evolutionary optimization procedure can, of course, not only be applied 
to the design of new pedestrian facilities, but also to a reduction of existing bottlenecks, when
suitable modifications are implemented. 

\section{Future Directions}

In this contribution, we have presented a multi-agent approach to pedestrian and crowd dynamics. 
Despite the great effort required, pedestrian interactions can be well quantified by video tracking.
Compared to other social interactions they turn out to be quite simple. Nevertheless, they cause a 
surprisingly large variety  of self-organized patterns and short-lived social phenomena, 
where coordination or cooperation emerges spontaneously. For this reason, they are 
interesting to study, particularly as one can expect new insights into coordination mechanisms of social
beings beyond the scope of classical game theory. 
Examples for observed self-organization phenomena in normal situations
are lane formation, stripe formation, oscillations and intermittent clogging effects at bottlenecks, and the
evolution of behavioral conventions (such as the preference of the right-hand side in continental Europe).
Under extreme conditions (high densities or panic), however, coordination may break down, giving
rise to ``freezing-by-heating'' or ``faster-is-slower effects'', stop-and-go waves or ``crowd turbulence''.

Similar observations as in pedestrian crowds are made in other social systems and settings.
Therefore, we expect that realistic models of pedestrian dynamics will
also promote the understanding
of opinion formation and other kinds of collective behaviors. The hope is that,
based on the discovered elementary mechanisms of emergence and 
self-organization, one can eventually also obtain a better understanding of 
the constituting principles of more complex social systems.
At least the same underlying factors are found in many social systems: non-linear interactions of individuals, 
time-dependence, heterogeneity, stochasticity, competition for scarce resources (here: space and time), 
decision-making, and learning. Future work will certainly also address issues of perception, anticipation, 
and communication.

\subsection*{Acknowledgments}

The authors are grateful for partial financial support by the German Research Foundation (research projects He 2789/7-1, 8-1) 
and by the ``Cooperative Center for Communication Networks Data Analysis'', a NAP project sponsored by
the Hungarian National Office of Research and Technology under grant
No.\ KCKHA005.

\end{document}